\documentclass[11pt,a4paper]{article}

\usepackage[T1]{fontenc}
\usepackage{eurosym}
\usepackage{booktabs}
\usepackage{multirow}
\usepackage{sectsty}
\usepackage{graphicx}
\usepackage{amsmath}
\usepackage{amssymb}
\usepackage{subcaption}
\usepackage{caption}
\usepackage{todonotes}
\usepackage{wrapfig}
\usepackage{hyperref}
\hypersetup{colorlinks=true,linkcolor=blue,urlcolor=blue}

\usepackage{fancyhdr}
\usepackage[left=2cm,top=1.5cm,bottom=1.5cm,right=2cm,includefoot,includehead,headheight=13.6pt]{geometry}
\usepackage{lastpage}
\usepackage{enumitem}
\usepackage{lipsum}
\usepackage{longtable}
\usepackage[toc]{multitoc}

\numberwithin{equation}{section}

\newcommand{\simas}[1]{\raisebox{-.1ex}{
            $\stackrel{\small{#1}}{\sim}$}}

\newcommand{\ssfill}{\xleaders\hbox to 0.35em{\scriptsize.}\hfill}

\makeatletter
\newcommand{\oset}[3][0ex]{%
	\mathrel{\mathop{#3}\limits^{
			\vbox to#1{\kern-2\ex@
				\hbox{$\scriptstyle#2$}\vss}}}}
\makeatother

\newcommand*{\cventry}[7][.25em]{
  \noindent\begin{tabular*}{\textwidth}{l@{\extracolsep{\fill}}r}%
	  {\bfseries #4} & {\bfseries #5} \\%
	  {\itshape #3\ifthenelse{\equal{#6}{}}{}{, #6}} & {\itshape #2}\\%
  \end{tabular*}%
  \ifx&#7&%
    \else{\\\vbox{\small#7}}\fi%
  \par\addvspace{#1}}

\newcommand*{\hintfont}{\bfseries}
\newcommand*{\hintstyle}[1]{{\noindent\hintfont{#1}}}
\newcommand*{\cvitem}[3][.25em]{%
  \ifthenelse{\equal{#2}{}}{}{\hintstyle{#2}: }{#3}%
  \par\addvspace{#1}}

\let\OriginalQuotation\quotation
\renewcommand*{\quotation}{\OriginalQuotation\small\sf}

\newcommand{\comment}[1]{}
\newcommand{\be}{\begin{equation}}
\newcommand{\ee}{\end{equation}}
\newcommand{\ba}{\begin{array}}
\newcommand{\ea}{\end{array}}
\newcommand{\baa}{\begin{array}}
\newcommand{\eaa}{\end{array}}
\newcommand{\bea}{\begin{eqnarray}}
\newcommand{\eea}{\end{eqnarray}}
\newcommand{\half}{\frac{1}{2}}
\newcommand{\MS}{{\overline{\rm MS}}}
\newcommand{\tl}{\tilde l}
\newcommand{\tL}{\tilde L}
\newcommand{\mur}{\mu_{\rm had}}
\newcommand{\mupt}{\mu_{\rm pt}}

\newcommand{\lambdat}{ \lambda_{\rm TGF}}

\newcommand{\LTGF}{ \Lambda_{\rm TGF}}

\newcommand{\Tr}{ {\rm Tr}}
\newcommand{\bn}{\overline{n} }
\usepackage{array}
\newcolumntype{\Vert}{!{\vrule width 1pt}}

\begin{document}

\allsectionsfont{\sffamily}

\fancyhead{}
\fancyfoot{}

\fancyhead[CO]{{}}
\fancyhead[LO]{{}}
\fancyhead[RO]{{}}
\fancyfoot[R]{\thepage\, / {\color[rgb]{0.6,0.,0}\pageref{LastPage}}}
\renewcommand{\headrulewidth}{0pt}

\newcommand*{\titleGP}{\begingroup 
\thispagestyle{empty}

\begin{flushright}
  IFT-UAM/CSIC-21-77 \\
  IFIC/21-25
\end{flushright}

{\centering 

\begin{center}
{\LARGE {\sf Memory efficient finite volume schemes with twisted boundary conditions}} \\[0.2\baselineskip]
\end{center}

}
\begin{center}
  {\large Eduardo~I.~Bribi\'an$^a$, Jorge~Dasilva~Gol\'an$^a$$^b$, Margarita~Garc\'{\i}a~P\'erez$^a$ and Alberto~Ramos$^c$}
\end{center}
\vspace{0.2cm}
\begin{center}
  $^a${Instituto de F\'{\i}sica Te\'orica UAM-CSIC, Nicol\' as Cabrera 13-15,
    Campus de Cantoblanco.\\
    28049, Madrid, Spain}\\
  $^b${Departamento de F\'{\i}sica Te\'orica, m\'odulo 15, Universidad Aut\'onoma de Madrid, Cantoblanco.\\
    28049, Madrid, Spain}\\
  $^c${Instituto de F\'{\i}sica Corpuscular (IFIC), CSIC-Universitat
    de Valencia.\\
    46071, Valencia, Spain}
\end{center}

\vspace{1cm}
\begin{center}
  \large{\sf Abstract}
\end{center}
\rule{\textwidth}{0.4pt}
\noindent
In this paper we explore a finite volume renormalization scheme that
combines three main ingredients: a coupling based on the gradient
flow, the use of twisted boundary conditions and a particular
asymmetric geometry, that for $SU(N)$ gauge theories consists on a
hypercubic box of size $l^2 \times (Nl)^2$, a choice motivated by the
study of volume independence in large $N$ gauge theories.  We argue
that this scheme has several advantages that make it particularly
suited for precision determinations of the strong coupling,
among them translational invariance, an analytic expansion in the
coupling and a reduced memory footprint with respect to standard
simulations on symmetric lattices, allowing for a more efficient use
of current GPU clusters.  We test this scheme numerically with a
determination of the $\Lambda$ parameter in the $SU(3)$ pure gauge
theory. We show that the use of an asymmetric geometry has no
significant impact in the size of scaling violations, obtaining a
value $\Lambda_\MS \sqrt{8 t_0} =0.603(17)$ in good agreement with the
existing literature. The role of topology freezing, that is relevant
for the determination of the coupling in this particular scheme and
for large $N$ applications, is discussed in detail.

\\\rule{\textwidth}{0.4pt}\\[\baselineskip] 

\tableofcontents

\newpage
\endgroup}

\begingroup 
\thispagestyle{empty}

\begin{flushright}
  IFT-UAM/CSIC-21-77 \\
  IFIC/21-25
\end{flushright}

{\centering 

\begin{center}
{\LARGE {\sf Memory efficient finite volume schemes with twisted boundary conditions}} \\[0.2\baselineskip]
\end{center}

}
\begin{center}
  {\large Eduardo~I.~Bribi\'an$^a$, Jorge~Dasilva~Gol\'an$^a$$^b$, Margarita~Garc\'{\i}a~P\'erez$^a$ and Alberto~Ramos$^c$}
\end{center}
\vspace{0.2cm}
\begin{center}
  $^a${Instituto de F\'{\i}sica Te\'orica UAM-CSIC, Nicol\' as Cabrera 13-15,
    Campus de Cantoblanco.\\
    28049, Madrid, Spain}\\
  $^b${Departamento de F\'{\i}sica Te\'orica, m\'odulo 15, Universidad Aut\'onoma de Madrid, Cantoblanco.\\
    28049, Madrid, Spain}\\
  $^c${Instituto de F\'{\i}sica Corpuscular (IFIC), CSIC-Universitat
    de Valencia.\\
    46071, Valencia, Spain}
\end{center}

\vspace{1cm}
\begin{center}
  \large{\sf Abstract}
\end{center}
\rule{\textwidth}{0.4pt}
\noindent
In this paper we explore a finite volume renormalization scheme that
combines three main ingredients: a coupling based on the gradient
flow, the use of twisted boundary conditions and a particular
asymmetric geometry, that for $SU(N)$ gauge theories consists on a
hypercubic box of size $l^2 \times (Nl)^2$, a choice motivated by the
study of volume independence in large $N$ gauge theories.  We argue
that this scheme has several advantages that make it particularly
suited for precision determinations of the strong coupling,
among them translational invariance, an analytic expansion in the
coupling and a reduced memory footprint with respect to standard
simulations on symmetric lattices, allowing for a more efficient use
of current GPU clusters.  We test this scheme numerically with a
determination of the $\Lambda$ parameter in the $SU(3)$ pure gauge
theory. We show that the use of an asymmetric geometry has no
significant impact in the size of scaling violations, obtaining a
value $\Lambda_\MS \sqrt{8 t_0} =0.603(17)$ in good agreement with the
existing literature. The role of topology freezing, that is relevant
for the determination of the coupling in this particular scheme and
for large $N$ applications, is discussed in detail.

\\\rule{\textwidth}{0.4pt}\\[\baselineskip] 

\tableofcontents

\newpage
\endgroup

\section{Introduction}
\label{sec:introduction}

Finite volume renormalization schemes~\cite{Luscher:1991wu} are a powerful tool to
investigate asymptotically free theories (see~\cite{Luscher:1998pe,Sommer:1997xw} for
a comprehensive review). 
Asymptotic freedom~\cite{Politzer:1973fx, Gross:1973id} predicts that
at high energies the coupling is weak, and perturbation theory becomes
a reliable tool to make precise predictions. 
In Yang Mills (YM) theories or QCD, the approach to this perturbative
regime is logarithmic. 
Due to this logarithmic running of the coupling, the strongly coupled
regime and the perturbative regime are separated by large energy scales. 
Accommodating these disparate scales on a single lattice simulations
is challenging, and requires compromises (see~\cite{DelDebbio:2021ryq}
for a review on the consequences in the context of the extraction of
the strong coupling). 
Finite volume renormalization schemes solve this multiscale problem by
linking the physical volume of the system with the renormalization scale. 
A single lattice simulation only allows to study a limited range of
scales, but by using a recursive procedure called finite size scaling,
simulations with different physical sizes allow to cover 
large energy scales. Using this technique, the running coupling has
been determined in QCD with $N_{\rm f} = 2$~,
$3$~\cite{Aoki:2009tf, DallaBrida:2016kgh, Bruno:2017gxd} and
$4$~\cite{Tekin:2010mm} flavors as well as in the pure gauge
theory~\cite{Luscher:1993gh, Ishikawa:2017xam, DallaBrida:2019wur}.  

In this work we will explore numerically a particular finite volume
renormalization scheme, first introduced
in~\cite{Bribian:2019ybc,Bribian:2020xfs,IbanezBribian:2019zef} based
on three ingredients. First, we choose a non-perturbative coupling based on the
gradient
flow~\cite{Narayanan:2006rf,Lohmayer:2011si,Luscher:2009eq,Luscher:2010iy} 
(see also~\cite{Ramos:2015dla} for a review).  
Second, we choose twisted boundary conditions~\cite{tHooft:1979rtg,tHooft:1980kjq,tHooft:1981sps} for
the gauge fields. 
Finally, we make use of a particular geometry, namely, for the case of
$SU(N)$ gauge theories we choose a hyper-cubic box of dimensions
$l^2\times \tilde l^2$ with $\tilde l = lN$. 

The logic behind these choices is rooted in the study of the large
$N$ limit of YM theories and in the ideas of volume
reduction with twisted boundary
conditions~\cite{GonzalezArroyo:1982hz,GonzalezArroyo:1982ub,GonzalezArroyo:2010ss} 
(see~\cite{Perez:2014sqa, GarciaPerez:2020gnf} for recent
reviews). With our choice of geometry ($l^2\times(Nl)^2$) and our
particular choice of twisted boundary conditions, perturbation theory
indicates that the effective volume of the torus is $(Nl)^4$. 
This equivalence between group degrees of freedom and spatial
degrees of freedom (that becomes exact in the large $N$ limit), lies
behind the ideas of volume reduction and non-commutative
geometry. 

In this work we will will argue that this particular choice of scheme
has several advantages.  
The running coupling in the pure gauge theory has
recently seen a revived interest in relation of the quest for
precision in the LHC. 
It has been shown that precise values of the $\Lambda$ parameter of
the pure gauge theory can be used, via a non-perturbative matching
between QCD and the pure gauge theory using heavy
quarks~\cite{DallaBrida:2019mqg}, into a precise value for the strong
coupling, a key quantity for phenomenology in high energy physics. 
Due to the fact that the pure gauge theory is numerically more
tractable than QCD this approach is probably the best option to
substantially reduce the uncertainty in the strong coupling. 
Still, very precise determinations of the running
coupling in the pure gauge theory are needed, that poses its own
challenges. 
Compared with the more customary finite volume
renormalization schemes based on Schr\"odinger
Functional (SF)~\cite{Luscher:1992an, Fritzsch:2013je} or mixed open-SF boundary
conditions~\cite{Luscher:2014kea}, twisted boundary
conditions~\cite{Ramos:2014kla,Keegan:2015lva} preserve the invariance
under translations and are therefore free of the linear $\mathcal
O(a)$ cutoff effects present in other schemes. 
The use of periodic boundary
conditions~~\cite{Fodor:2012td,Fodor:2012qh} lead to coupling
definitions that are non-analytic, making cumbersome the extraction of
the $\Lambda$ parameter. This is not the case for twisted boundary conditions, which are 
particularly suited for perturbative calculations.
Moreover, the relation between the $\Lambda$ parameter in this
scheme and in the $\overline{\rm MS}$ scheme is known
analytically~\cite{Bribian:2019ybc}.
Finally, due to the use of an asymmetrical torus, the memory footprint required
to simulate this scheme in a computer is reduced. 
The computing power stays roughly the same because with this geometry the
observables show a larger variance~\cite{Nada:2020jay}, but this
scheme allows for a better usage of current GPU clusters.

In this work we will determine the $\Lambda$ parameter in the pure
gauge theory. 
We will check that the use of this particular geometry has no
significant impact in the size of scaling violations, and will
explore in detail the role of topology
freezing~\cite{DelDebbio:2004xh, Fritzsch:2013yxa}. 
In summary we expect that this particular scheme will play an important
role in future precise determinations of the strong coupling, and in
the investigation of the large $N$ limit of gauge theories.

Let us now comment a bit on our choice of coupling and geometry. 
Definitions of the gauge coupling based on the gradient flow are
nowadays customary since they can be computed on
the lattice to high precision.
The gradient flow is a renormalization procedure  based on replacing
the original gauge fields $A_\mu(x)$ by a new set of smooth, flow time dependent,
fields $B_\mu(x,t)$, in terms
of which gauge invariant composite observables are automatically renormalized quantities for $t>0$~\cite{Luscher:2011bx}.
These new fields are driven by the so-called flow time equations:
\be
\partial_t B_\mu (x, t) = D_\nu G_{\nu \mu} (x, t), \quad B_\mu (x, t = 0) = A_\mu (x),
\ee
where $D_\mu$ and $ G_{\mu \nu}$ stand for the covariant derivative and field strength tensor of the flow fields.
The effect of the flow is to smooth the gauge fields on a sphere of radius $\sqrt{8t}$.
Renormalized couplings based on the gradient flow can thus be
trivially introduced,  as one only needs to find dimensionless observables
that depend on a scale. 
Starting from $E(t)$, the energy density of the flow fields:
\be
E (t) = \half  \Tr \left (G_{\mu \nu} (x, t)G_{\mu \nu} (x, t)\right)\, ,
\ee
we can use the flow time $t$, with dimensions of length squared, to
construct the dimensionless combination $\langle t^2 E(t)\rangle $. 
Nevertheless, this dimensionless combination depends on a scale $\mu =
1/\sqrt{8t}$, which allows to define a renormalized 't Hooft coupling ($\lambda= g^2 N$) 
through the relation:
\be
\lambda (\mu) = {\cal N} \left<t^2E\left(t\right)\right>\Big |_{\sqrt{8t}=\mu^{-1}},
\ee
where ${\cal N}$ stands for a normalization constant introduced to
ensure that $\lambda(\mu) = \lambda_{\overline{\rm MS} }(\mu) + ...$
at leading order in perturbation theory.  
In the context of finite-size scaling, it is customary to relate the
renormalization scale to the size of the finite volume in which the 
gauge theory lives. In the standard set-up, on a torus of size $l^4$,
one fixes the flow time via the relation $\sqrt{8t}=c l$, with $c$
standing for an arbitrary, smaller that 1,  constant defining the
scheme.  

Concerning the geometry, the Twisted Gradient Flow (TGF) scheme is defined by introducing $SU(N)$ Yang-Mills theories on an 
asymmetrical hypercubic box of size $l^2 \times (N l)^2$.
The plane with two short directions, taken to be the (1,2) plane, is endowed with twisted boundary conditions, i.e.
\be
A_\mu (x + l \hat \nu ) = \Gamma_\nu A_\mu (x) \Gamma_\nu^\dagger , \text{ for } \nu  = 1,2.
\ee
In the remaining directions the gauge field is periodic with period $\tl=N l$.
The twist matrices $\Gamma_\nu$, for $\nu$  = 1, 2, satisfy the consistency condition:
\be
\Gamma_1 \Gamma_2 = Z_{12} \Gamma_2 \Gamma_1
\ee
with
\be
Z_{12}  = e^{i  \frac{2 \pi k}{N}},
\ee
with $k$ and $N$ coprime integers.
For the concrete case of $SU(3)$ studied in this paper, the choice of
twist is unique and given by $k=1$.  
However, for arbitrary $N$ the  value of the twist $k$ has to be scaled with the range of the gauge group and 
is best selected so as to avoid the appearance of tachyonic instabilities in the 
large $N$ limit~\cite{GonzalezArroyo:2010ss,Chamizo:2016msz}. 

As already mentioned, perturbation theory indicates that finite volume effects for this choice of twist are
controlled  by $\tl$ ~\cite{Perez:2018afi,Perez:2013dra,Bribian:2019ybc}. 
Under this assumption, it becomes natural to use the effective size $\tl$, rather that $l$, to set the scale of 
the running coupling.
In addition, and following the prescription introduced in \cite{Fritzsch:2013yxa}, the
coupling will be defined within the sector of configurations
with zero topological charge. This choice aims at circumventing the
problem of topology freezing on the lattice that will be discussed in
some detail in sec.~\ref{sec:topology}. With this, the coupling is defined as: 
\begin{equation}
        \lambdat(\mu) =\frac{ 128 \pi^ 2 t^2 }{3 N  {\cal A}(\pi c^2)} \frac{\langle E\left(t\right)\delta_Q\rangle}{\langle \delta_Q\rangle} \Big |_{_{\sqrt{8t}=c \tl = \mu^{-1}}} \, ,
        \label{eq:lambdat}
\end{equation}
where $\delta_Q$ stands for a $\delta$-function that restricts the path integral to configurations with zero topological charge $Q$, and 
\be
{\cal A} (x) = x^2 \theta_3^2(0,ix) \left (\theta_3^2(0,ix) - \theta_3^2 (0,ix N^2) \right )
\, ,
\label{eq:calA}
\ee
where $\theta_3(0,ix)$ stands for the Jacobi $\theta_3$ function:
\be
\theta_3 ( z ,i x)  = \sum_{m \in \mathbb{Z}} \exp \left \{ - \pi x m^2 + 2 \pi i m z \right \} \equiv 
\frac{1}{\sqrt{x}} \sum_{m \in \mathbb{Z}} \exp \left \{ - \pi \frac{(m-z)^2}{x} \right \}  \, .
\label{eq:theta3}
\ee
Throughout this work we make use of the coupling scheme corresponding to $c=0.3$.

The paper is organized as follows. Section~\ref{sec:lambda-parameter} summarizes the strategy to determine the 
$\Lambda$ parameter in terms of some typical hadronic scale. Section~\ref{sec:numerical-results} compiles our numerical results. In subsection~\ref{s_setup} we introduce our numerical set-up and the lattice definition of the TGF 
coupling. Subsection~\ref{s.step} discusses the determination of the step scaling function and the continuum extrapolation. In subsection~\ref{sec:LTGF} we determine the $\Lambda$ parameter in the TGF scheme in terms of a hadronic scale $\mur$, and discuss the matching with the SF and $\MS$ schemes.  
Our final result for $\Lambda_\MS \sqrt{8t_0} $ is presented in subsection~\ref{s_Lambda}. Finally, 
section~\ref{sec:topology} discusses the role of topology freezing. We conclude in section~\ref{sec:conclusions}
with a summary of results. A few appendices are included providing raw data for our lattice results, details on the matching to the SF scheme and a discussion of the dilute gas approximation used in sec.~\ref{sec:topology}. 


\section{The Lambda parameter}
\label{sec:lambda-parameter}
The starting point for the determination of the $\Lambda$ parameter is
the RG equation in a certain scheme (labeled ${\rm s}$)
\be
\beta_{\rm s}(\lambda_{\rm s}) = \frac{{\rm d} \lambda_{\rm s}}{{\rm d} \log(\mu^2)}\,.
\label{eq:betaf} \ee
The $\beta$-function has a perturbative expansion
\begin{equation}
  \beta_{\rm s}(\lambda_{\rm s}) \simas{\lambda_{\rm s}\to 0} -\lambda_{\rm s}^2(b_0 + b_1\lambda_{\rm s}) + \dots\,.
\end{equation}
with the first two
coefficients $(4
\pi)^2 b_0= 11/3$ and $(4 \pi)^4 b_1 = 34/3 $ being universal (i.e. 
scheme independent).
The first order differential equation~\eqref{eq:betaf} can be integrated exactly 
\bea \frac{\Lambda_{\rm s}}{\mu} &=&\left[b_0
\lambda_{\rm s}(\mu)\right]^\frac{-b_1}{2{b_0}^2}\exp\left(\frac{-1}{2b_0
\lambda_{\rm s}(\mu)}\right) \nonumber\\
&\times&\exp\left[-\int_0^{\lambda_{\rm s}(\mu)} dx
\left(\frac{1}{2 \beta_s(x)}+\frac{1}{2 b_0 x^2}-\frac{b_1}{2
b_0^2x}\right)\right]\,,
\label{eq:LTGF} \eea
where $\Lambda_{\rm s}$ is an integration constant (the $\Lambda$-parameter). As
such, it is a renormalization group invariant (i.e. 
scale independent), but depends on the choice of renormalization
scheme. This scheme dependence
can be computed exactly via a one-loop computation. In
particular, the quotient of $\Lambda$ parameters in the $\MS$ scheme
and the TGF one (see section~\ref{sec:introduction} for a precise
definition of the latter) is  
given by~\cite{Bribian:2019ybc}
\be
\log\left(\frac{\LTGF}{\Lambda_\MS}\right)=\frac{3}{22}\left(\frac{11}{3}\gamma_E+\frac{52}{9}-3\log3+C_1(c)\right),
\label{eq:LTGF_LMS} \ee where $C_1(c=0.3)=0.508(4)$ for gauge group
$SU(3)$ and twist $k=1$.

In order to determine the $\Lambda$ parameter, we use the exact
relation eq.~\eqref{eq:LTGF} to write
\be \frac{\Lambda_{\rm s}}{\mur} = \frac{\Lambda_{\rm s} }{\mupt} \times
\exp\left[-\int_{\lambda_{\rm s}(\mupt)}^{\lambda_{\rm s}(\mur)} \frac{dx}{2 \beta_{\rm s}(x)}
\right]\,,
\label{eq:LTGF2} \ee
where $\mur$ is a typical hadronic scale and $\mupt \gg \mur$ is a
high energy scale where perturbation theory is applicable. 
Lattice field theory allows to determine non-perturbatively the
running of the coupling in some suitable schemes (like the TGF used
here), and therefore the second factor of the
right hand side of eq.~\eqref{eq:LTGF2}. In particular the technique
of finite size scaling~\cite{Luscher:1991wu} allows very precise determinations. 
We postpone a detailed discussion for the next section, and focus here
on the evaluation of the first factor, that requires the use of
perturbation theory. If we define 
\be I_{\rm s}^{(n)}(\lambda) = \exp \left
\{-\int_0^{\lambda} dx \left(\frac{1}{2 \beta_{\rm s}^{(n)}(x)}+\frac{1}{2 b_0
x^2}-\frac{b_1}{2 b_0^2x}\right)\right \}\,,
\label{eq:Ioflambda} \ee
where $\beta_{\rm s}^{(n)}(x)$ is the $n-$loop $\beta$-function, it is
easy to see that (cf. eq.\eqref{eq:LTGF})
\be
\frac{\Lambda_{\rm s}}{\mupt} \simas{\lambda_s(\mupt)\to 0} \left(b_0 \lambda_s(\mupt)\right)^\frac{-b_1}{2{b_0}^2}
\exp\left(\frac{-1}{2b_0 \lambda_s(\mupt)}\right)
I_{\rm s}^{(n)}\left (\lambda_s(\mupt)\right) + \mathcal O\left (\lambda_s^{n-1}(\mupt)\right) \,.
\label{eq:LTGF3} 
\ee
The corrections decrease as powers of $\lambda_s(\mupt)$, and therefore
logarithmically (i.e. 
slowly) with the scale $\mupt$. 
The extraction of the $\Lambda_{\rm s}$ parameter has to
be performed by taking the limit 
\begin{equation}
  \frac{\Lambda_{\rm s}}{\mur} = \lim_{\lambda_s(\mupt) \to 0}
  \left(b_0 \lambda_s(\mupt)\right)^\frac{-b_1}{2{b_0}^2}
\exp\left(\frac{-1}{2b_0 \lambda_s(\mupt)} \right) I_s^{(n)}\left(\lambda_s(\mupt)\right)
  \times
\exp\left[-\int_{\lambda_s(\mupt)}^{\lambda_s(\mur)} \frac{dx}{2 \beta_{\rm s}(x)}
\right]\,.
\label{eq:lambdalim}
\end{equation}
Controlling the limit of eq.~(\ref{eq:lambdalim}) requires simulating
large energy scales with our lattice simulations. 
Finite size scaling allows to do this by linking the renormalization
scale with the physical size of the lattice simulation (see
section~\ref{sec:introduction}). The change of the coupling when the
renormalization scale changes by a factor two
\begin{equation}
  \label{eq:sigmadef}
  \sigma(u) = \lambda_s(\mu/2)\Big|_{\lambda_s(\mu) = u}\,,
\end{equation}
is called the step scaling function, and can be obtained via lattice
simulations by first tuning the value of the bare coupling $\lambda_0$ such
that $\lambda_s = u$ at several values of the lattice size $\tilde L =
\tl/a$, and then determining the value of the coupling in a lattice twice
as large at the same values of the bare coupling $\lambda_0$. 
This procedure results in several values of the lattice step scaling
function $\Sigma(u,\tL)$, that have to be extrapolated to the
continuum 
\begin{equation}
  \sigma(u) = \lim_{1/\tilde L \to 0}\Sigma(u, \tilde L) \,.
\end{equation}

Since
\be
\int_{\lambda_s(\mu)}^{\lambda_s(\mu/2)}\frac{dx}{\beta_s(x)} =
\int_{u}^{\sigma(u)}\frac{dx}{\beta_s(x)} =
-2\log 2\,,
\label{eq:logs} \ee
the step scaling function allows to determine precisely the last
factor in Eq.~(\ref{eq:lambdalim}). Once the function $\sigma(u)$ is
known, one defines $u_0 = \lambda_s(\mur)$, and recursively determines
\begin{equation}
  \label{eq:useq}
u_k = \sigma^{-1}(u_{k-1})\,. 
\end{equation}
Each application of the step scaling function changes the
renormalization scale by a factor $2$, so that $u_{k} =
\lambda_s(2^k\mur)$ and
\begin{equation}
  \int_{u_k}^{u_0}\frac{{\rm d} x}{\beta_s(x)} = -2k\log 2\,.
\end{equation}
After a reasonable number of steps ($\mathcal O(10)$), large energy
scales have been achieved, where perturbative corrections are expected
to be small and one can explore the limit of Eq.~(\ref{eq:lambdalim})
by taking $\lambda_s(\mupt) = u_k$ and $\mu_{\rm pt} = 2^k\mur  \gg \mur$.

Crucially, large energy scales can be covered using this procedure. 
In the next section we discuss the numerical determination of the step
scaling function in the TGF set-up.


\section{Numerical results}
\label{sec:numerical-results}
\subsection{Numerical set-up}

\label{s_setup}

We simulate pure $SU(3)$ Yang-Mills theory on asymmetric lattices of
size $L^2 \times \tL^2$, with $\tL=N L\equiv 3 L$, corresponding to a torus in
the continuum of size $l^2 \times \tl^2$, with $\tl = a \tL$ and $a$
the lattice spacing.  Twisted boundary conditions are implemented by
taking as lattice action
\be S_W(U) = 3 b \sum_{n} \sum_{\mu \ne \nu}
\text{Tr }[ 1 - Z^*_{\mu\nu}(n) P_{\mu\nu}(n) ]\,,
\ee
with plaquettes
$P_{\mu\nu}(n)$ given in terms of the $SU(3)$ link variables
$U_\mu(n)$ as: \be P_{\mu\nu}(n)=U_\mu(n) U_\nu(n+\hat{\mu})
U^\dagger_\mu(n+\hat{\nu}) U^\dagger_\nu(n), \ee and where $b=
1/\lambda_0 $ stands for the inverse of the lattice
bare 't Hooft coupling, and $Z_{\mu\nu}(n)$ is set to one for all
plaquettes except for the ones with coordinates $x_1 = x_2 = 1$, where:
\be Z_{12} = Z^*_{21} = \exp \{ i 2 \pi / 3 \} \,.  \ee We use a
hybrid over-relaxation algorithm~\cite{Wolff:1992nq}, that combines a single
heat-bath sweep (HB)~\cite{Creutz:1980zw, Fabricius:1984wp}, followed
by $\tilde{L}$ over-relaxation sweeps (OR)~\cite{Creutz:1987xi}. 
Since the measurement of flow quantities are computationally
expensive, we perform $\tilde{L}$ hybrid sweeps between measurements,
which produces uncorrelated measurements basically for all our
coupling values (nevertheless, the small autocorrelations are still
taken into account). The simulated values of
the bare coupling and lattice sizes as well as the total number of
configurations attained in each case is reported in
table~\ref{tableA.1} in appendix~\ref{sec:esembles}.  We have used
lattices with $\tL= 12$, 18, 24, 36 and 48, in a range of values of $b
\in [0.33,1.12]$.

In this study, we have used the so-called Wilson flow
discretization of the continuum flow equation, and
evaluated the energy density by means of the clover discretization
given by: 
\bea 
E_{\rm cl}(t) &=& 
\frac{1}{2}\text{Tr}\left[G^{\rm cl}_{\mu\nu}(n,t) G^{\rm cl}_{\mu\nu}(n,t)\right],
\eea 
where we defined, denoting $U_{-\mu}(n)=U^\dagger_\mu(n-\mu)$:
\begin{align} 
G^{\rm cl}_{\mu\nu}(n,t) \nonumber &= -\frac{i}{8} \{
Z^*_{\mu\nu}(n) P_{\mu\nu}(n,t) + Z^*_{\mu\nu}(n-\hat{\nu})
P_{-\nu\mu}(n,t) \nonumber \\ & + Z^*_{\mu\nu}(n-\hat{\mu})
P_{\nu-\mu}(n,t) + Z^*_{\mu\nu}(n-\hat{\mu}-\hat{\nu})
P_{-\mu-\nu}(n,t) -c.c. \} ,
\label{eq.gmunu}
\end{align} with $P_{\mu\nu}(n,t)$ the plaquette evaluated in terms of
the flowed links at flow time $t$.

It is well known that simulations at very fine lattice spacings loose
ergodicity~\cite{DelDebbio:2004xh}. 
In many of our simulations this is not problematic, since they
are performed in small physical volumes, where the contributions of
the non-trivial topological sectors are highly suppressed. But at
physical volumes $\tl \sim 1$ fm, critical slowing down can severely
affect the results of our studies~\cite{Fritzsch:2013yxa}. 
For this reason, we define the coupling in the zero topological sector
using:
\be 
\lambdat(\mu) ={\cal N} (c, \tilde L) \frac{\langle
t^2E_{\rm cl}\left(t\right)\hat \delta_Q\rangle}{3 \langle \hat
\delta_Q\rangle} \Big |_{_{\sqrt{8t}=c \tl = \mu^{-1}}} \, ,
\label{eq:lambdat_latt} 
\ee 
where the topological charge $Q$ on the lattice is determined through the 
expression: 
\be 
Q= \frac{1}{16 \pi^2} \sum_n \Tr \left \{
G^{\rm cl}_{\mu\nu}(n,t) \widetilde G^{\rm cl}_{\mu\nu}(n,t) \right \} 
\label{eq:qlatt}
\ee 
evaluated at flow time $\sqrt{8 t}= c \tl$.  Since the topological charge given by
this expression is not an integer, we have set: 
\be 
\hat \delta_Q =
\left \{ \begin{array}{cc} 1 & \text{if } |Q| < 0.5 \\ 0 &
\text{otherwise} \\
 \end{array} \right. . 
\ee 
In addition, and in order to eliminate the
leading order lattice artefacts in perturbation theory, the lattice 
normalization factor has been used: 
\bea 
{\cal N}^{-1}(c, \tilde L) &=&
\frac{c^4}{128}\sum_{\mu\neq\nu} \sum_{q}^{'} e^{-\frac{1}{4}c^2\tilde{L}^2\hat{q}^2}\frac{1}{\hat{q}^2}\sin^2(q_\nu)\cos^2(q_\mu/2),
\label{latt_norm} 
\eea 
where $\hat q_\mu =2\sin(q_\mu/2)$ stands for
the lattice momentum, with $q_\mu = 2 \pi n_\mu /\tL$, with $n_\mu = 0, \cdots, \tL-1$,
and with the prime in the sum denoting the exclusion of momenta with
both components in the twisted plane satisfying $\tL q_i \propto 2 N \pi = 6\pi$.

The values for the TGF coupling measured for $c=0.3$ are collected in
table~\ref{tableB.1} in appendix~\ref{sec:raw-data}.  At a given value
of $b$, the coupling is measured on two lattices with extents given by
$\tL$ and $2\tL$, which allows to determine the lattice step scaling function
$\Sigma(u, \tilde L)$ for $\tilde L = 12, 18, 24$.

\subsection{Determination of the continuum step scaling function}

\label{s.step}

Once the lattice step scaling function $\Sigma(u, \tilde L)$ is
determined, one needs to extrapolate it to the continuum. 
In our set-up, leading cutoff effects are $\mathcal O(a^2)$. 
Even though this leading behavior receives logarithmic
corrections~\cite{Husung:2019ytz},
our range of lattice sizes span a factor two in lattice spacings, so 
we expect these logarithmic corrections to be small. Since the aim of
this work is not to provide a very precise result, but else to explore the viability of the 
TGF scheme, we will rely on simple linear extrapolations.
If data at fixed values of the coupling $u_i\, (i=1,\dots,N_{\rm
  points})$ and their corresponding values of the lattice step scaling
function $\Sigma_i(\tilde L) \equiv \Sigma(u_i, \tilde L)$ exist
for different lattice sizes $\tilde L$, we can extrapolate them as:
\begin{equation}
  \Sigma_i(\tilde L) = \sigma_i + \frac{m_i}{\tilde L^2}\,.
\end{equation}
Alternatively, we have tried to extrapolate
\begin{equation}
  \frac{1}{\Sigma_i(\tilde L)} = \frac{1}{\sigma_i} + \frac{m'_i}{\tilde L^2}\,.
\label{eq.Sigma_latt}
\end{equation}
The difference between the two approaches is clearly an $1/\tilde
L^4$ term, and in our checks it turns out to be negligible. 
These extrapolations at fixed values of $u_i$ result in the
corresponding values $\sigma_i = \sigma(u_i)$ of the step scaling
function. This  approach is described in detail in section~\ref{s.ubyuf}.

It is also convenient to find out a suitable parametrization for the
continuum function $\sigma(u)$, in order to obtain the sequence of
couplings eq.~\eqref{eq:useq}.
Reasonable parametrizations for the functional form $\sigma(u)$ can be
found by using perturbation theory as a guide. In particular, we will
use the simple parametrization
\begin{equation}
\frac{1}{\sigma(u)} = \frac{1}{u} -2b_0 \log 2 - 2b_1 u \log 2 +
\sum_{k=2}^{N_c} p_k u^k
\label{eq.invsu} 
\end{equation}
to fit our data. 
Note that although this functional form imposes the correct perturbative
behavior (for $u\to 0$), after determining the fit parameters $p_k$
using our non perturbative data, we expect it provides a
description of the non-perturbative step scaling function $\sigma(u)$ in our range
of couplings. 

Alternatively, being $\sigma(u)$ a one-to-one function in our
domain of interest, one can use the inverse relation to parameterize
the inverse step scaling function $\sigma^{-1}$ entering in eq.~\eqref{eq:useq}, i.e. 
\begin{equation}
  \frac{1}{u(\sigma)} = \frac{1}{\sigma}
+2b_0 \log 2 + 2b_1 \sigma \log 2 + \sum_{k=2}^{N_c} \hat p_k
\sigma^k .
\label{eq.invss} 
\end{equation}
The two parametrizations should give equivalent results. 

Finally we can combine both, the continuum limit and the
parametrization of the step scaling function, in a single global fit. 
This approach does not require the coupling to be perfectly tuned to
constant values of $u$, and amounts to fit the data for the lattice
step scaling function to the functional form
\begin{equation}
  \label{eq:globalfit}
\frac{1}{\Sigma(u, \tilde L)} = \frac{1}{u} -2b_0 \log 2 - 2b_1 u \log 2 +
\sum_{k=2}^{N_c} p_k u^k +
\left(  \sum_{k = 0}^{N_l} \rho_k u^k \right) \times \frac{1}{\tilde L^2},
\end{equation}
and similarly for the parametrization of the lattice inverse function
\begin{equation}
{\cal U}(x, \tL) = \Sigma^{-1}(x, \tilde L)\,.
\end{equation}
Details of this global approach are presented in section~\ref{s.globalf}.

\begin{table}
  \centering
  \begin{tabular}{ccccc}
  \toprule
    $u_{\rm tg}$ & $\Sigma(u_{\rm tg}, 12)$ & $\Sigma(u_{\rm tg}, 18)$ & $\Sigma(u_{\rm tg}, 24)$  & $\sigma(u_{\rm tg})$\\
    \midrule
1.8425	&	1.9503(34)	&	1.9558(37)	&	1.9577(61)	&	1.9602(55)	\\
2.0568	&	2.1967(38)	&	2.2002(46)	&	2.2113(68)	&	2.2092(65)	\\
2.3270	&	2.5031(45)	&	2.5153(50)	&	2.5147(85)	&	2.5224(75)	\\
2.6834	&	2.9271(52)	&	2.9519(59)	&	2.9411(99)	&	2.9607(88)	\\
3.1746	&	3.5160(66)	&	3.5278(71)	&	3.554(11)	&	3.555(10)	\\
3.4945	&	3.9068(75)	&	3.9502(81)	&	3.955(11)	&	3.978(11)	\\
3.8881	&	4.4249(88)	&	4.475(12)	&	4.477(12)	&	4.501(14)	\\
4.3739	&	5.088(11)	&	5.113(19)	&	5.133(17)	&	5.143(20)	\\
5.0260	&	5.997(13)	&	6.058(18)	&	6.095(20)	&	6.120(22)	\\
5.9170	&	7.337(17)	&	7.453(24)	&	7.471(27)	&	7.527(29)	\\
7.2325	&	9.570(26)	&	9.758(31)	&	9.732(37)	&	9.836(39)	\\
7.9431	&	10.933(31)	&	11.202(41)	&	11.257(48)	&	11.386(51)	\\
8.8322	&	12.975(44)	&	13.223(54)	&	13.339(73)	&	13.434(71)	\\
9.9945	&	16.148(63)	&	16.575(78)	&	16.585(96)	&	16.81(10)	\\
13.91650&	32.01(21)	&	33.21(27)	&	34.00(25)	&	34.52(29)	\\
\bottomrule
  \end{tabular}
  \caption{Values of the lattice step scaling function
    $\Sigma(u_{\rm tg},\tilde L)$ at fixed targeted values of the coupling $u_{\rm tg}$ and for
    different values of $\tilde L$. We also report on the continuum
    extrapolation $\sigma(u_{\rm tg})$.}
  \label{tab:ubyudata}
\end{table}

\subsubsection{Continuum limit on a $u$-by-$u$ basis}
\label{s.ubyuf}

We have tuned the bare coupling $b$ on the $\tilde{L}=12$, $18$ and
$24$ lattices to attain a few targeted values of the renormalized
coupling $u$ (see table~\ref{table_u_cont} in appendix~\ref{ap.extrap}).  
Once this is achieved, the renormalized coupling is
measured on the double size lattices at the same value of the bare
coupling. Since the tuning is not perfect, there is a small mismatch 
in the tuning of the target couplings that can be easily corrected for 
by slightly shifting the resulting couplings to a constant value of $u$. 

By using the one-loop perturbative relation
\be
\frac{1}{\Sigma(u, \tilde L)} - \frac{1}{u} = \text{constant}\,,
\ee
the values of the lattice step scaling function $\Sigma(u,\tilde L)$
can be shifted to a target value of the coupling $u_{\rm tg}$ using
the relation
\be
\Sigma(u_{\rm tg}, \tilde L) =
\Sigma(u,\tilde L) + \frac{\Sigma^2(u,\tilde L)}{u^2}(u-u_{\rm tg})\,.
\ee
We note that our data set is already very well tuned,
and therefore the additional shift introduced by this procedure 
is well below the statistical accuracy.

The same one-loop relation can be used to propagate the error of $u$ into an
error of the step scaling function.
Being precise, we add in quadratures
\begin{equation}
  \delta \Sigma(u,\tilde L) = \frac{\Sigma^2(u, \tilde L)}{u^2}\delta u
\end{equation}
to the statistical error of $\Sigma(u,\tilde L)$.
This procedure leads to a set of data at constant value of the
coupling $u$ for different lattice spacings.
The result can be seen in table~\ref{tab:ubyudata}, while the raw data
is available in table~\ref{table_u_cont} in appendix~\ref{ap.extrap}.

\begin{figure}[t] \centering
\begin{subfigure}{.5\textwidth} \centering
\includegraphics[width=\linewidth]{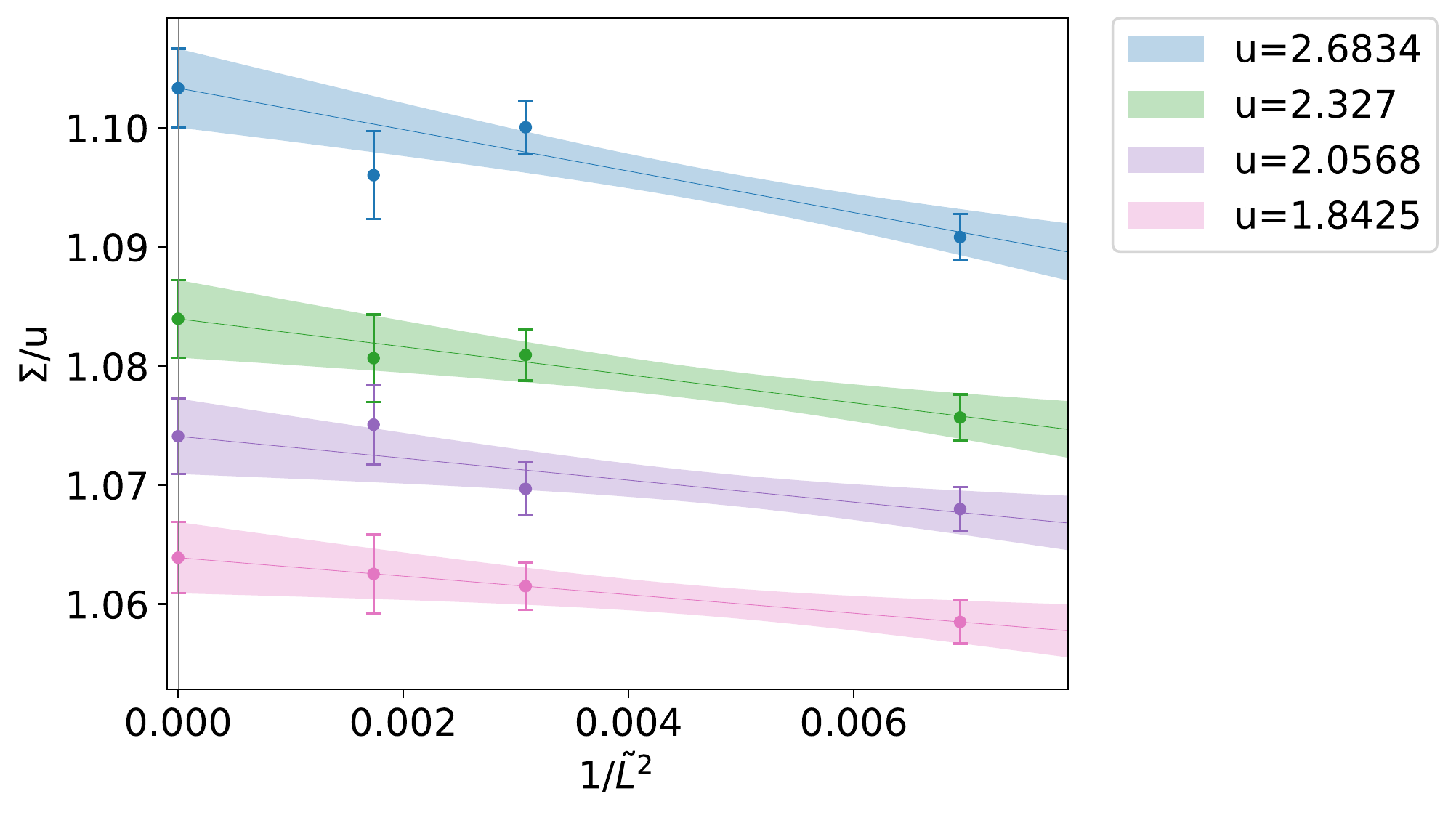}
\end{subfigure}%
\begin{subfigure}{.5\textwidth} \centering
\includegraphics[width=\linewidth]{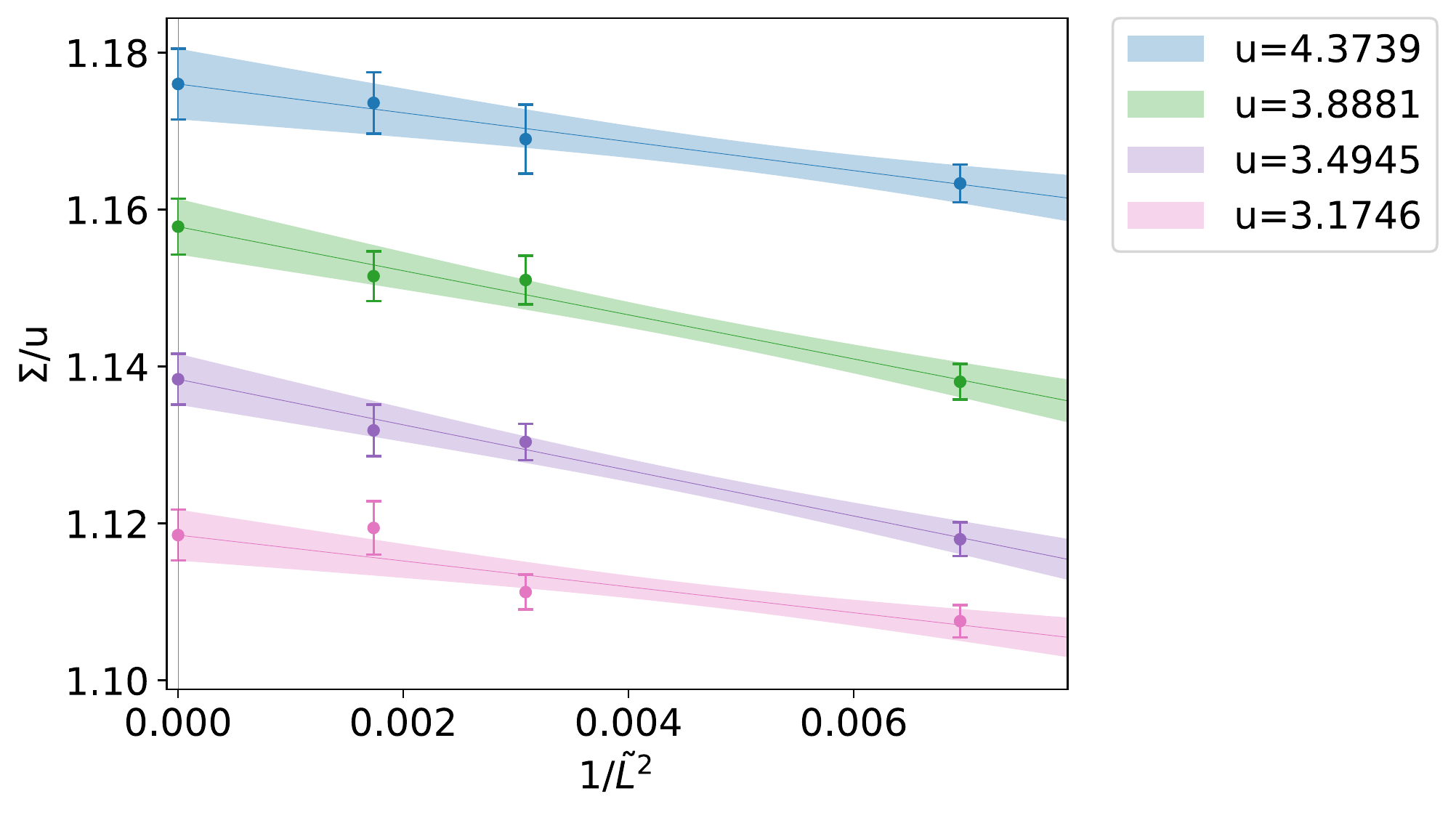}
\end{subfigure}
\begin{subfigure}{.5\textwidth} \centering
\includegraphics[width=\linewidth]{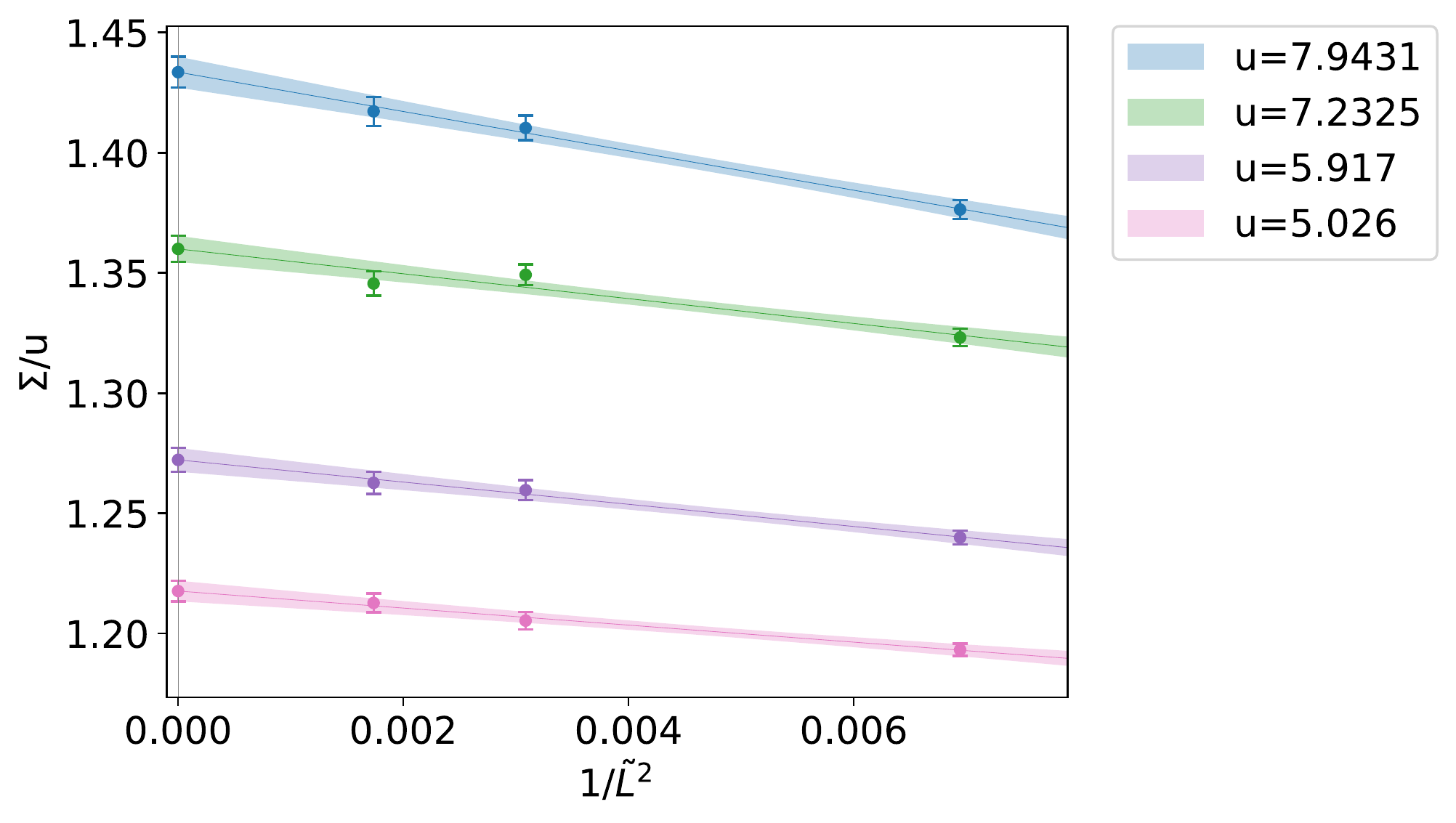}
\end{subfigure}%
\begin{subfigure}{.5\textwidth} \centering
\includegraphics[width=\linewidth]{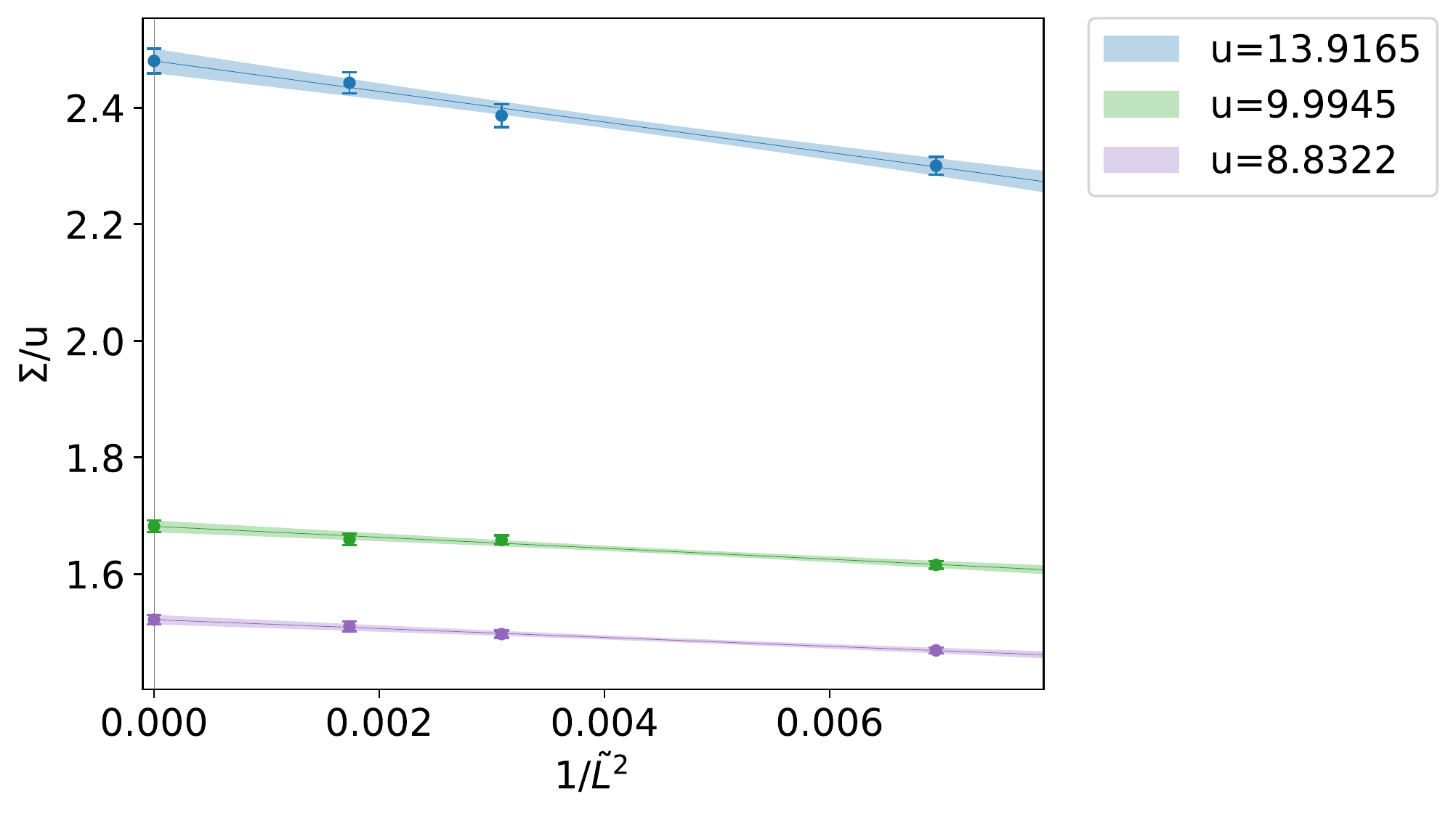}
\end{subfigure}
\caption{We display the continuum extrapolation of $\Sigma(u)/u$,
performed at a few selected values of $u$.}
\label{Figure_extrapl_fixed_u}
\end{figure}

The continuum limit of the step scaling function is obtained by
performing a linear extrapolation in $1/\tL^2$,
as illustrated in fig.~\ref{Figure_extrapl_fixed_u}. The continuum
values of $\sigma(u)$ resulting from these fits are given in
table~\ref{tab:ubyudata}.
The advantage of this procedure vs the global fit presented below is that no functional
dependence in $u$ is assumed for the terms representing the lattice
artefacts (cf. eqs.~\eqref{eq.Sigma_latt} and~\eqref{eq:globalfit}).

The resulting pairs $(u,\sigma)$ are then fitted to the series expansions
given by eqs.~\eqref{eq.invsu} or \eqref{eq.invss}, allowing to
parameterize the corresponding step scaling functions in the range of
couplings under consideration.  For completitude we give the
parameters of the fit to eq.~\eqref{eq.invss} with $N_c=4$ and fitting range $\sigma \in
[1.5, 17]$:
\be
\hat p_2 = -2.86020692\times 10^{-5},\, \,   \hat p_3 = 1.94775566\times 10^{-6} ,\, \,    \hat p_4 = -4.21570896 \times 10^{-8}, 
\ee
and their covariance:
\be
\begin{pmatrix}
\, 9.62610704\times 10^{-10}  & -  1.48907565 \times 10^{-10} & \, 5.49335753\times 10^{-12}	\\
-  1.48907565\times 10^{-10}  & \, 2.37337095 \times 10^{-11} & -  8.93899741\times 10^{-13}	\\
\, 5.49335753\times 10^{-12}  & -  8.93899741 \times 10^{-13} & \, 3.41858249\times 10^{-14} 
\end{pmatrix} \, .
\ee

\begin{figure}[h] \centering
\begin{subfigure}{.5\textwidth} \centering
\includegraphics[width=\linewidth]{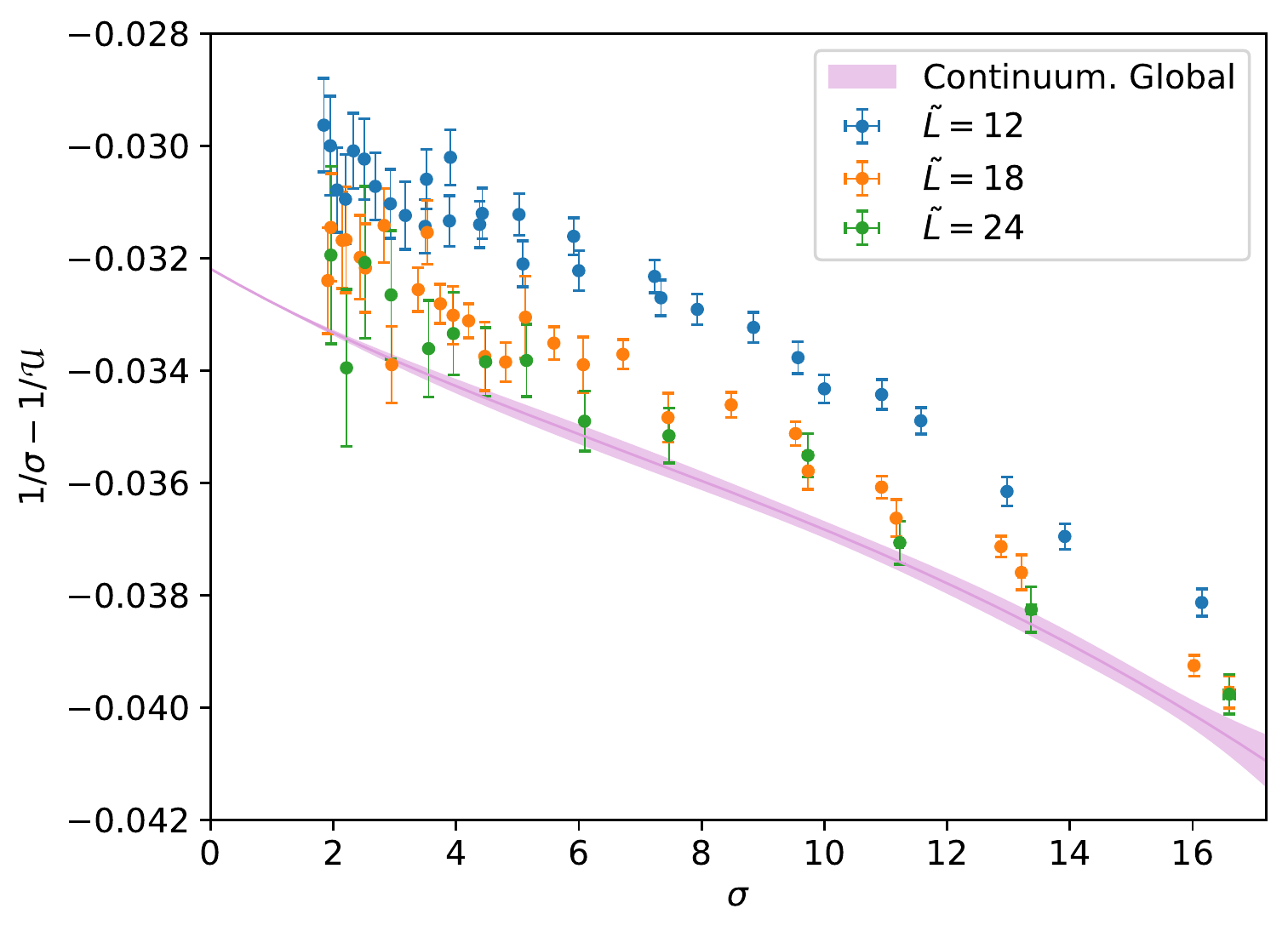}
\caption{Raw data for $1/\sigma - 1/{\cal U}(\sigma, \tL)$ vs
$\sigma$.}
\label{Figure_extrapla}
\end{subfigure}%
\begin{subfigure}{.5\textwidth} \centering
\includegraphics[width=\linewidth]{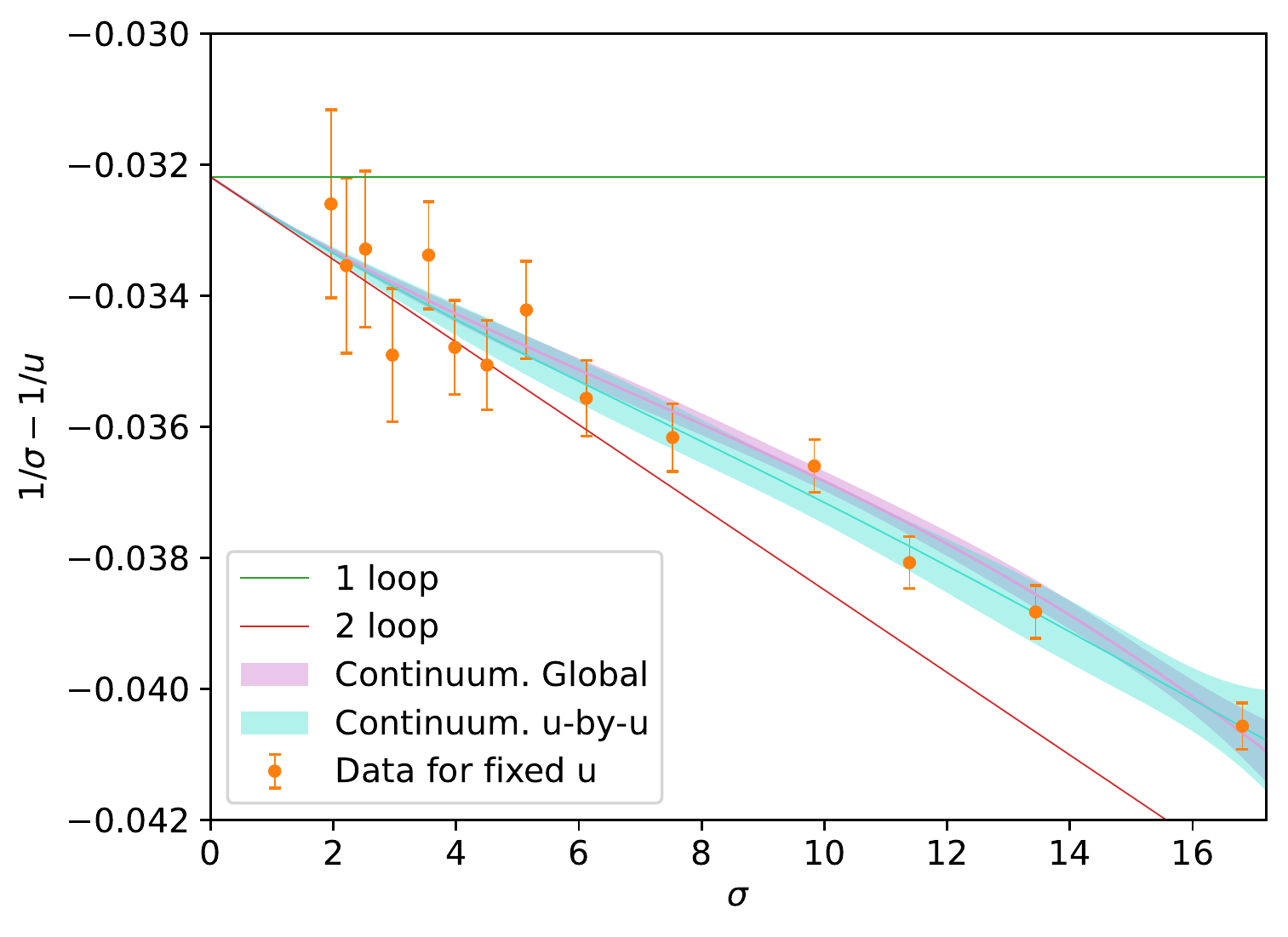}
\caption{ Continuum extrapolation of $1/\sigma - 1/u$ vs $\sigma$.}
\label{Figure_extraplb}
\end{subfigure}
\caption{(a) Raw data for the inverse step scaling function
compared to the continuum extrapolation obtained
from the global fit, cf. eqs.~\eqref{eq.fitglob},~\eqref{eq.fitglobcov}.  
(b) We display the comparison between various continuum extrapolations of
$1/\sigma - 1/u$ vs $\sigma$. The orange data points correspond to the determination
at fixed values of $u$ given in table~\ref{tab:ubyudata}.
The continuum extrapolations derived
from the $u$-by-$u$ and the global fits described in the text, correspond respectively to the blue and pink
bands, while the continuous lines indicate the perturbative
predictions at order one-loop and two-loops. }
\label{Figure_extrapl}
\end{figure}

\subsubsection{Continuum limit from a global fit}
\label{s.globalf}

As an alternative to the $u$-by-$u$ fit, the continuum limit can be obtained from a combined
parametrization of the continuum step scaling function and a continuum
extrapolation -- see e.g. eq.~(\ref{eq:globalfit}). 

From the last section, we know that the continuum step scaling
function is well parametrized using three parameters in addition to the two universal terms (i.e. $N_{\rm c}=4$). 
Cutoff effects are well described as long as $N_{\rm l} \ge 3$, with little differences
in the continuum results even if the number of parameters is 
increased up to $N_{\rm l} = 8$. 
The raw data used in these fits is available in appendix~\ref{sec:raw-data}. 
We have fitted the data for $1/{\cal U}(\sigma, \tL)$ vs $\sigma$ in the range 
$\sigma \in[1.5, 17]$ with the particular choice $N_{\rm c} = N_{\rm l} = 4$;
the resulting continuum fit parameters are given by:
\be
\hat p_2 = -3.50622141\times 10^{-5},\, \,   \hat p_3 = 2.14311070 \times 10^{-6} ,\, \,    \hat p_4 = -2.97994910 \times 10^{-8},
\label{eq.fitglob}
\ee
with covariance:
\be
\begin{pmatrix}
\, 2.87681194\times 10^{-10}  & -  4.50822649 \times 10^{-11} & \, 1.69327683 \times 10^{-12}    \\
-  4.50822649 \times 10^{-11} & \, 7.22888354 \times 10^{-12} & -  2.76073298 \times 10^{-13}    \\
\, 1.69327683\times 10^{-12}  & -  2.76073298 \times 10^{-13} & \, 1.06806168 \times 10^{-14}
\end{pmatrix}
\, .
\label{eq.fitglobcov}
\ee
The result of this fitting procedure is displayed in fig.~\ref{Figure_extrapla} where the raw data for 
$1/{\cal U}(\sigma, \tL)$ is represented vs $\sigma$ and compared with the continuum extrapolation resulting from
this global fit which is given by the pink band displayed in the plot. 

We finally present a comparison between the various continuum
extrapolations in fig.~\ref{Figure_extraplb}. The orange points come from the
direct extrapolation at fixed targeted values of $u$ given in
table~\ref{tab:ubyudata}. The blue and
pink bands correspond respectively to the $u$-by-$u$ and
global fits described previously, showing a good agreement in all the
range of fitted values. For comparison, the perturbative predictions at one and
two-loop order are also shown in the figure. 

\subsection{Determination of $\Lambda_{\rm TGF}/\mur$}
\label{sec:LTGF}
The determination of $\Lambda_{\rm TGF}/\mur$ relies on the use of
eq.~\eqref{eq:lambdalim}, evaluated in terms of a sequence of couplings
connecting $\lambda(\mur)$ and $\lambda(\mupt)$. 
We have performed this sequence in two steps: an exact step connecting the coupling at scales
$\mur$ and $2 \mur$, followed by a step-scaling sequence, starting at
$\lambda(2\mur)$, to be determined by the continuum step scaling
functions obtained in the previous sections. 
Let us now describe these steps in detail.

\begin{table}[t] \centering
  \begin{tabular}{lllllll}
    \toprule
    $k$ & \multicolumn{2}{c}{TGF} &\multicolumn{2}{c}{MS} &\multicolumn{2}{c}{SF} \\
    \cmidrule(lr){2-3} \cmidrule(lr){4-5}     \cmidrule(lr){6-7}
     & $\lambda(2^{k+1}\mu_{\rm had})$ & $\Lambda_{\overline{\rm MS} }/(2\mu_{\rm had})$
     & $\lambda_{\overline{\rm MS} }(2^{k+1}\mu_{\rm had})$ & $\Lambda_{\overline{\rm MS} }/(2\mu_{\rm had})$
     & $\lambda_{\rm SF}(0.9\times 2^{k+1}\mu_{\rm had})$ & $\Lambda_{\overline{\rm MS} }/(2\mu_{\rm had})$ \\
0  &13.9164955 & 0.18900(10) & 7.6539(49) & 0.11959(18) & 7.35(93) & 0.207(67)\\
1  &9.035(13)  & 0.19019(56) & 6.3953(76) & 0.14936(52) & 5.934(36) & 0.2263(44)\\
2  &6.7989(88) & 0.19271(71)& 5.3041(72) & 0.16265(80) & 4.864(11) & 0.2212(20)\\
3  &5.4781(96) & 0.1948(12) & 4.5077(69) & 0.1708(11) & 4.1391(87) & 0.2180(22)\\
4  &4.5986(99) & 0.1965(18) & 3.9148(70) & 0.1763(16) & 3.6138(81) & 0.2167(27)\\
5  &3.9683(96) & 0.1978(24) & 3.4591(69) & 0.1804(21) & 3.2127(75) & 0.2164(32)\\
6  &3.4933(89) & 0.1989(29) & 3.0987(66) & 0.1836(25) & 2.8950(69) & 0.2168(36)\\
7  &3.1220(81) & 0.1998(34) & 2.8068(61) & 0.1861(29) & 2.6364(63) & 0.2173(41)\\
8  &2.8234(74) & 0.2005(38) & 2.5656(57) & 0.1881(33) & - & -\\
9  &2.5778(67) & 0.2011(41) & 2.3630(52) & 0.1898(37) & - & -\\
10 &2.3723(60) & 0.2017(44) & 2.1903(48) & 0.1912(40) & - & -\\
11 &2.1976(55) & 0.2021(47) & 2.0414(44) & 0.1925(42) & - & -\\
12 &2.0472(50) & 0.2026(50) & 1.9117(41) & 0.1935(45) & - & -\\
13 &1.9163(45) & 0.2029(52) & 1.7976(37) & 0.1945(47) & - & -\\
14 &1.8014(41) & 0.2032(54) & 1.6965(35) & 0.1953(49) & - & -\\
$\infty$ &0.0&       0.2082(83) & 0.0 &       0.2097(76) & 0.0 &       0.2175(61)\\
    \bottomrule
\end{tabular}
\caption{We list the couplings along the step scaling sequence
starting at $\lambda(2\mur)=13.9164955$, corresponding to the three
different computation strategies. The values of $\Lambda_s/(2\mur)$
are obtained by matching to perturbation theory at the corresponding
value of $\lambda_s(\mupt)$. The final result, quoted in the last line, is
obtained from a linear fit to the last 5 steps in the sequence (3 in the case of SF).}
\label{table_Lambda}
\end{table}

\begin{description}
\item[Determination of $\lambda(\mu_{\rm had})$] 
We fix the coupling $\lambda(2\mur)\equiv 13.9164955$
at values of the lattice spacing that correspond to $\tilde L =
12,18,24$. 
The corresponding values of the lattice step scaling function are given in table~\ref{tab:ubyudata}, and
extrapolate to a continuum value $\lambda(\mur)= 34.52(29)$. 

\item[Determination of $\Lambda_{\overline{\rm MS}}/(2\mu_{\rm had})$]
Using the fits of the continuum step scaling function, we determine a
sequence of couplings $\lambda_n = \lambda(2^{n+1}\mu_{\rm had})$. 
The differences between the global and the u-by-u fits in this step are
well below our statistical uncertainties. 
On the other hand, there are several possibilities to determine the
ratio $\Lambda_{\overline{\rm MS} }/(2\mu_{\rm had})$:
\begin{description}
\item[TGF:] One can use the universal coefficients of the
  $\beta$-function and the known ratio $\Lambda_{\overline{\rm MS}
  }/\Lambda_{\rm TGF}$ to determine $\Lambda_{\overline{\rm MS}
  }/(2\mu_{\rm had})$ (cf. 
  eq.~(\ref{eq:lambdalim})).
\item[MS:] One can use the value of the ratio  $\Lambda_{\overline{\rm MS}
  }/\Lambda_{\rm TGF}$ to determine the coupling in the $\overline{\rm
    MS} $ scheme as follows
  \begin{equation}
    \lambda_{\overline{\rm MS}}(2^{n+1}\mu_{\rm had})  =
    x + 2b_0 x^2 \log \left( \frac{\Lambda_{\overline{\rm MS}}
  }{\Lambda_{\rm TGF}} \right) + \mathcal O(x^3)\,
\qquad\left(x = \lambda\left (2^{n+1}\mu_{\rm had}\right)\right)\,.
\end{equation}
Then, using the known 5-loop $\beta$-function~\cite{Baikov:2016tgj} in the
$\overline{\rm MS} $ scheme, one can
determine the ratio $\Lambda_{\overline{\rm MS} }/(2\mu_{\rm had})$.
\item[SF:] Very precise data for the SF coupling is available in reference~\cite{DallaBrida:2019wur}. 
  These computations were performed at the same values of the bare
  coupling $b$ as our runs but on lattices of size $\tL /3$. 
  This allows to convert non-perturbatively the values of
  $\lambda(2^{n+1}\mu_{\rm had})$ to values of $\lambda_{\rm
  SF}(0.9\times 2^{n+1}\mu_{\rm had})$.
  Details of this matching are available in appendix~\ref{sec:matching-with-sf}, here it is
  enough to say that this matching is only possible in the region
  $9 \gtrsim \lambda \gtrsim 3$.

  After this non-perturbative conversion, the known 3-loop
  $\beta$-function in the SF scheme~\cite{Bode:1998hd, Bode:1999dn,
    Bode:1999sm} can be used to determine  the ratio
  $\Lambda_{\overline{\rm MS} }/(2\mu_{\rm had})$.

\end{description}

This procedure is similar to the approach taken
in~\cite{Nada:2020jay}, and we encourage the reader interested in the
details to consult that reference.

\end{description}

\begin{figure}
  \centering
  \includegraphics[width=\textwidth]{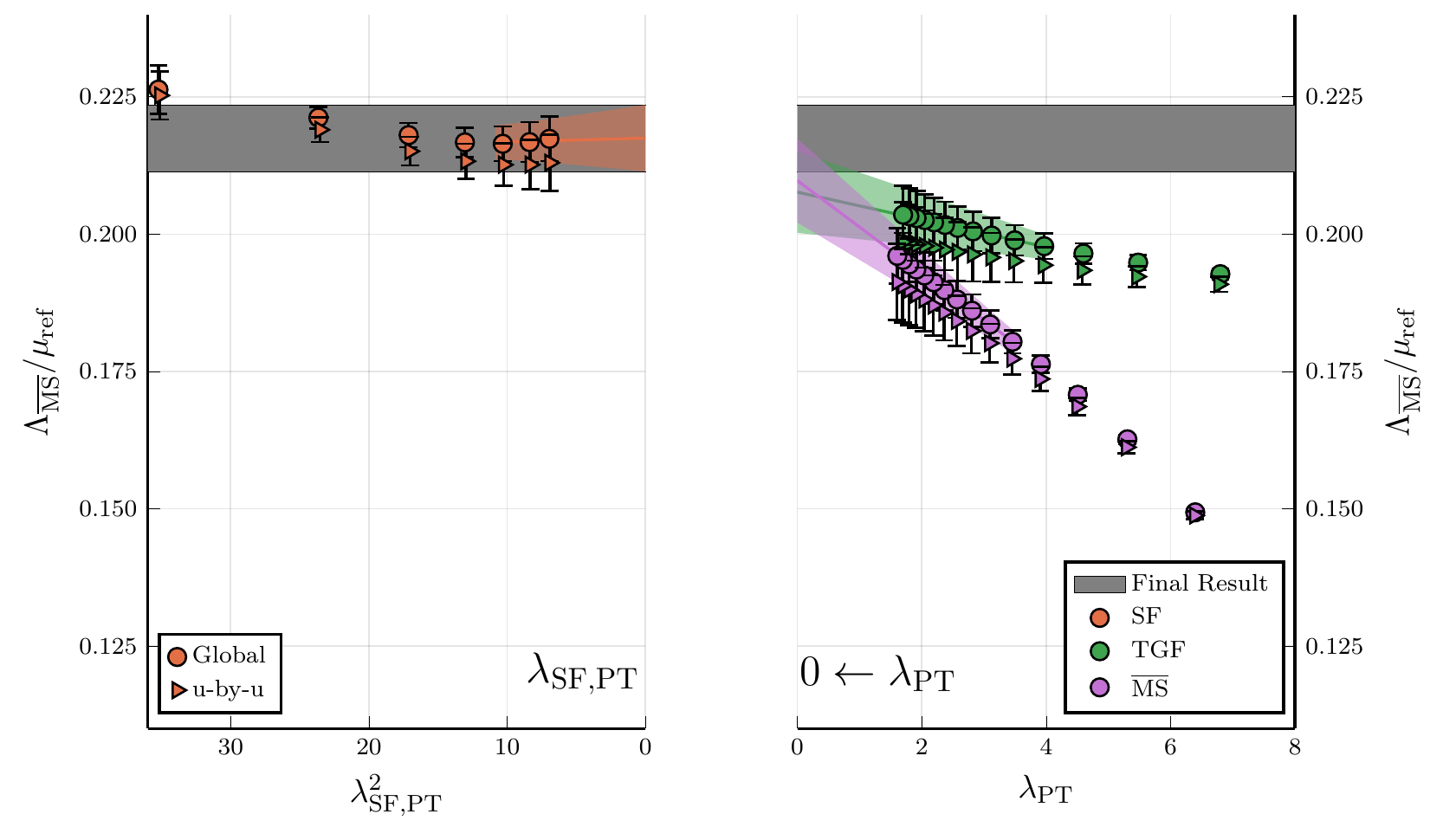}
  \caption{Ratio $\Lambda_{\overline{\rm MS} }/(2\mu_{\rm had})$ obtained through the different methods explained in the text as a function of the matching scale with perturbation theory ($\mu_{\rm ref} \equiv 2 \mur$). The grey band corresponds to the final result obtained from a linear fit to the 3 last steps in the SF sequence.}
  \label{fig:lseq}
\end{figure}

The results for the ratio $\Lambda_{\overline{\rm MS} }/(2\mu_{\rm had})$
using the different procedures described above can be seen in
table~\ref{table_Lambda} and in figure~\ref{fig:lseq}. 
Methods labeled TGF and MS carry a perturbative uncertainty $\mathcal
O(\lambda(2^{n+1}\mu_{\rm had}))$ (right panel of
figure~\ref{fig:lseq}), while the SF method carries an 
uncertainty  $\mathcal O(\lambda^2(2^{n+1}\mu_{\rm had}))$ (left
panel). Figure~\ref{fig:lseq} shows that these perturbative corrections are
significant in the methods labeled TGF and MS. Nevertheless the
picture shows a nice agreement between all analysis methods once the
limit $\lambda \to 0$ is taken. 

Finally, we quote as central value of the analysis the one labeled as SF, since it has the smallest perturbative 
corrections and is compatible within errors with the other two determinations. 
We report as final numerical value:
\begin{equation}
  \frac{ \Lambda_{\overline{\rm MS} }}{2\mu_{\rm had}} = 0.2175(61)\,.
\label{eq.Lambda_muhad}
\end{equation}

\subsection{Computation of $\Lambda_{\MS}\sqrt{ 8 t_0}$}
\label{s_Lambda}

\begin{figure}[t]
\centering
\includegraphics[width=0.6\textwidth]{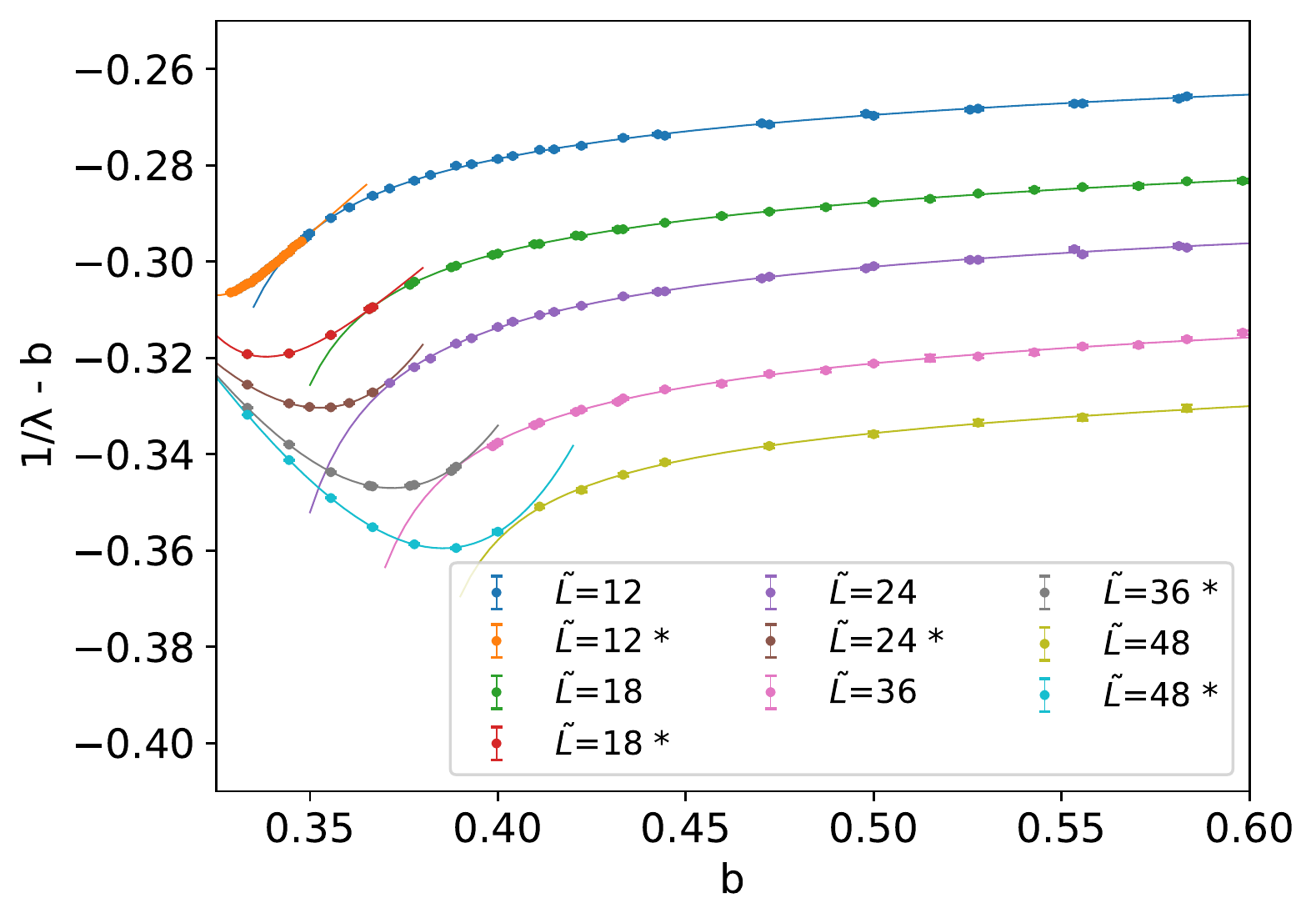}
\caption{ We display $1/\lambda- b $ as a function of $b$. The
lines represent Pad\'e fits of the form given in
eq.~\eqref{blambda}.} 
\label{plotblambda}
\end{figure}

\begin{table}
  \centering
  \begin{tabular}{ll|ll|ll}
    \toprule
    $18 \times b$&$t_0/a^2$&$18 \times b$&$t_0/a^2$&$18 \times b$&$r_0/a$ \\
    \midrule
5.9600 & 2.7854(62)  \(^\dagger\)      & 6.4200 & 11.241(23)  \(^\dagger\) & 5.7000 & 2.9220(90)  \(^{\dagger\dagger}\)\\  
5.9600 & 2.7875(53)  \(^\ddagger\)     &  6.4500 & 12.196(21) \(^\ddagger\)        & 5.8000 & 3.6730(50)  \(^{\dagger\dagger}\)\\  
6.0500 & 3.7834(47)  \(^\ddagger\)     &  6.5300 & 15.156(28) \(^\ddagger\)        & 5.9500 & 4.898(12)  \(^{\dagger\dagger}\)\\   
6.1000 & 4.4329(32)  \(^{\star}\)  &  6.6100 & 18.714(30)  \(^\ddagger\)     &  6.0700 & 6.033(17)  \(^{\dagger\dagger}\)\\   
6.1300 & 4.8641(85)  \(^\ddagger\)     &  6.6720 & 21.924(81) \(^{\star}\)   & 6.1000 & 6.345(13)  \(^{\star}\)\\       
6.1700 & 5.489(14) \(^\dagger\)       &   6.6900 & 23.089(48) \(^\ddagger\)        & 6.2000 & 7.380(26)  \(^{\dagger\dagger}\)\\   
6.2100 & 6.219(13)  \(^\ddagger\)     &   6.7700 & 28.494(66) \(^\ddagger\)        & 6.3400 & 9.029(77)  \(^{\star}\)\\       
6.2900 & 7.785(14)  \(^\ddagger\)     &   6.8500 & 34.819(84)  \(^\ddagger\)       & 6.4000 & 9.740(50)  \(^{\dagger\dagger}\)\\   
6.3400 & 9.002(31)  \(^{\star}\)  &  6.9000 & 39.41(15) \(^{\star}\)      & 6.5700 & 12.380(70)  \(^{\dagger\dagger}\)\\  
6.3400 & 9.034(29)  \(^{\star}\)  &   6.9300 & 42.82(11) \(^\ddagger\)           &6.5700 & 12.176(97)\(^{\mathsection}\) \\        
6.3700 & 9.755(19)  \(^\ddagger\)     &    7.0100 & 52.25(13)  \(^\ddagger\)      & 6.6720 & 14.103(92)  \(^{\star}\)\\        
6.4200 & 11.202(21)  \(^\ddagger\)    &           &                       &     6.6900   &     14.20(12)\(^{\mathsection}\) \\  
           & & & &  6.8100   &     16.54(12)\(^{\mathsection}\) \\
           & & & &  6.9000   & 18.93(15)  \(^{\star}\)  \\
           & & & &  6.9200   &     19.13(15)\(^{\mathsection}\) \\
    \bottomrule
  \end{tabular}
  \caption{Results for $t_0/a^2$ and $r_0/a$ for different values of $b$. 
  	       For the case of $t_0/a^2$ the relevant references are 
  	       $^\dagger$~\cite{Luscher:2010iy}, $^\ddagger$~\cite{Giusti:2018cmp}, 
  	       $^{\star}$~\cite{Knechtli:2017xgy}. 
  	       For $r_0/a$, instead, the results are from $^{\dagger\dagger}$~\cite{Guagnelli:1998ud} and $^{\star}$~\cite{Knechtli:2017xgy}. The data labelled $^{\mathsection}$ is obtained from the values of $r_c/a$ of~\cite{Necco:2001xg} together with $r_c/r_0=0.5133(24)$ which gives the quoted values of $r_0/a$.} 
  \label{tab:t0data}
\end{table}

In order to determine the dimensionless combination
$\Lambda_{\MS}\sqrt{ 8 t_0}$, we have to put together our previous results
for the ratio $\Lambda_{\overline{\rm MS} }/\mu_{\rm had}$ with a
determination of $\mu_{\rm had}\times \sqrt{8t_0}$. 
This quantity is
computed as the continuum extrapolation of 
\be 
\mu_{\rm had} \sqrt{8t_0} = \lim_{a \rightarrow 0} \, \left( a \mur \times
\frac{\sqrt{8 t_0}}{a}\right). 
\ee 
where $a \mur$ is determined by the lattice size and bare coupling 
for which the renormalized coupling takes the value $\lambda(\mur)=34.52(29)$ (see table~\ref{tab:ubyudata}).
Values for $\sqrt{8t_0}/a$ for our choice of Wilson gauge action are
available in the literature (see table~\ref{tab:t0data}).

Two steps were therefore required.  First, for each value of $\tL$ we determined
the values of the bare coupling leading to a renormalized coupling
$\lambda(\mur)$. For this purpose, we have fitted the dependence of the
coupling on $b$, at fixed $\tL$, with a Pad\'e fit of the form: 
\be 
b \lambda (b,\tL)=\frac{a_0+a_1b+b^2}{a_2+a_3b+b^2}.
\label{blambda} \ee 
Fits are done to $1/\lambda-b$, as illustrated in fig.~\ref{plotblambda}. 

\begin{table}[t]
\centering
\begin{tabular}{ c \Vert c  c  c  c  c}
$\tilde{L}$ & $18\times b_{\lambda=34.52}$ & $a/\sqrt{8t_0}$ & $a \mur\times \sqrt{8t_0}/a$ & $a/r_0$ & $a \mur \times r_0/a$ \\ [0.9ex]
\hline
12      &       6.0037(26)      &       0.19637(96)     &       1.4146(69)      &       0.1851(20)      &        1.501(16)      \\
18      &       6.2518(30)      &       0.13373(65)     &       1.3848(68)      &       0.1256(16)      &               1.474(18)       \\
24      &       6.4574(30)      &       0.10029(47)     &       1.3848(65)      &       0.0947(13)      &               1.467(21)       \\
36      &       6.7641(36)      &       0.06683(37)     &       1.3855(77)      &       0.0637(15)      &               1.453(34)       \\
48      &       6.9908(30)      &       0.05010(25)     &       1.3861(70)      &       0.0471(39)      &               1.47(12)        \\
\hline
$\infty$        &                       &       0               &       1.3860(62)      &          0            &  1.453(19)
\end{tabular}
\caption{Values of the bare coupling, lattice spacing and reference scales $\mur\sqrt{8t_0}$ and $\mur r_0$, corresponding to the renomalization
condition $\lambda(\mur)=34.52(29)$ -- see text.}
\label{t0fits}
\end{table}

\begin{figure}[t]
\centering
\begin{subfigure}{.45\textwidth} \centering
\includegraphics[width=\textwidth]{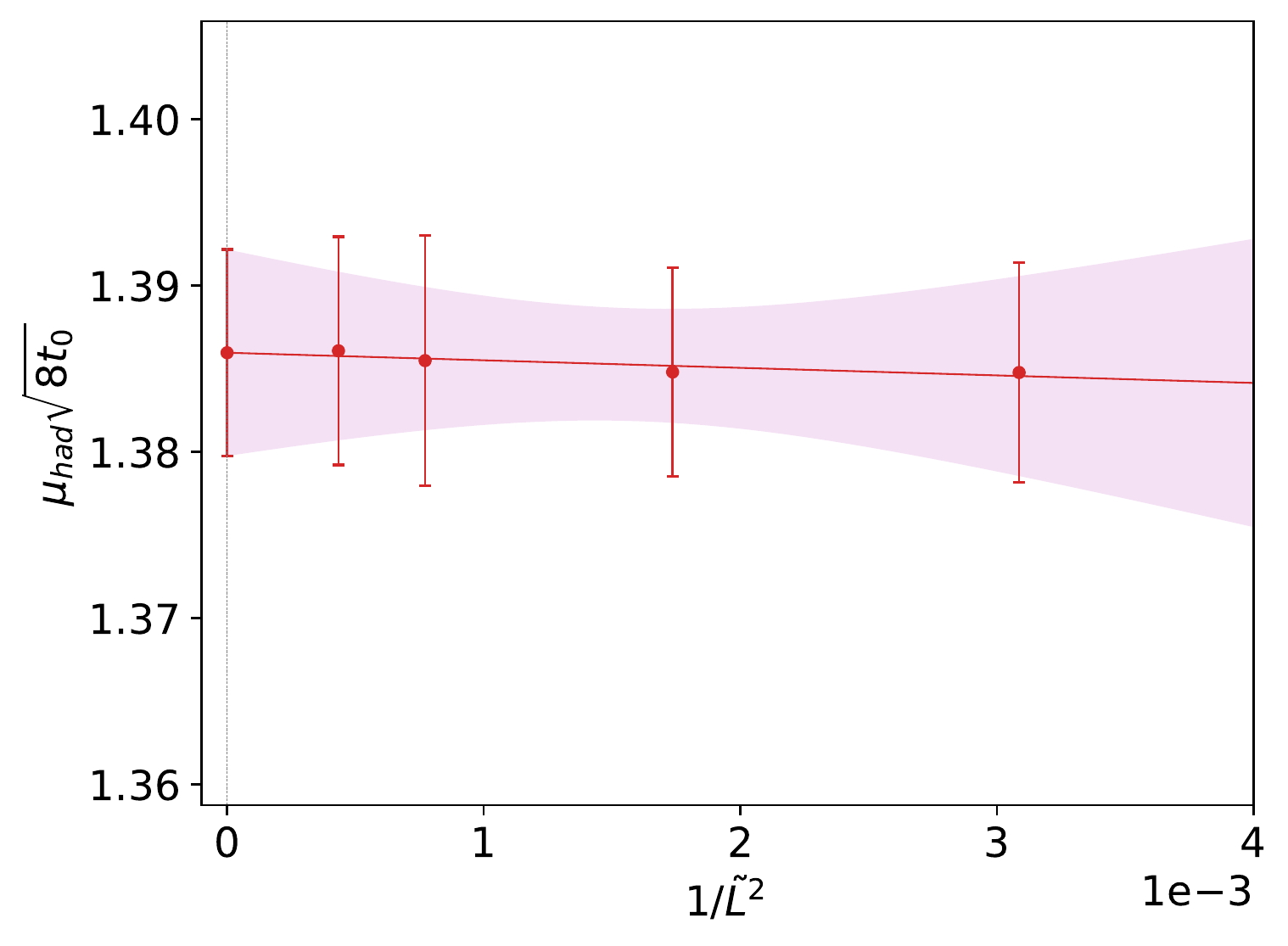}
\caption{$\mur \times \sqrt{8 t_0}$. }
\label{t0}
\end{subfigure}
\begin{subfigure}{.45\textwidth} \centering
\includegraphics[width=\textwidth]{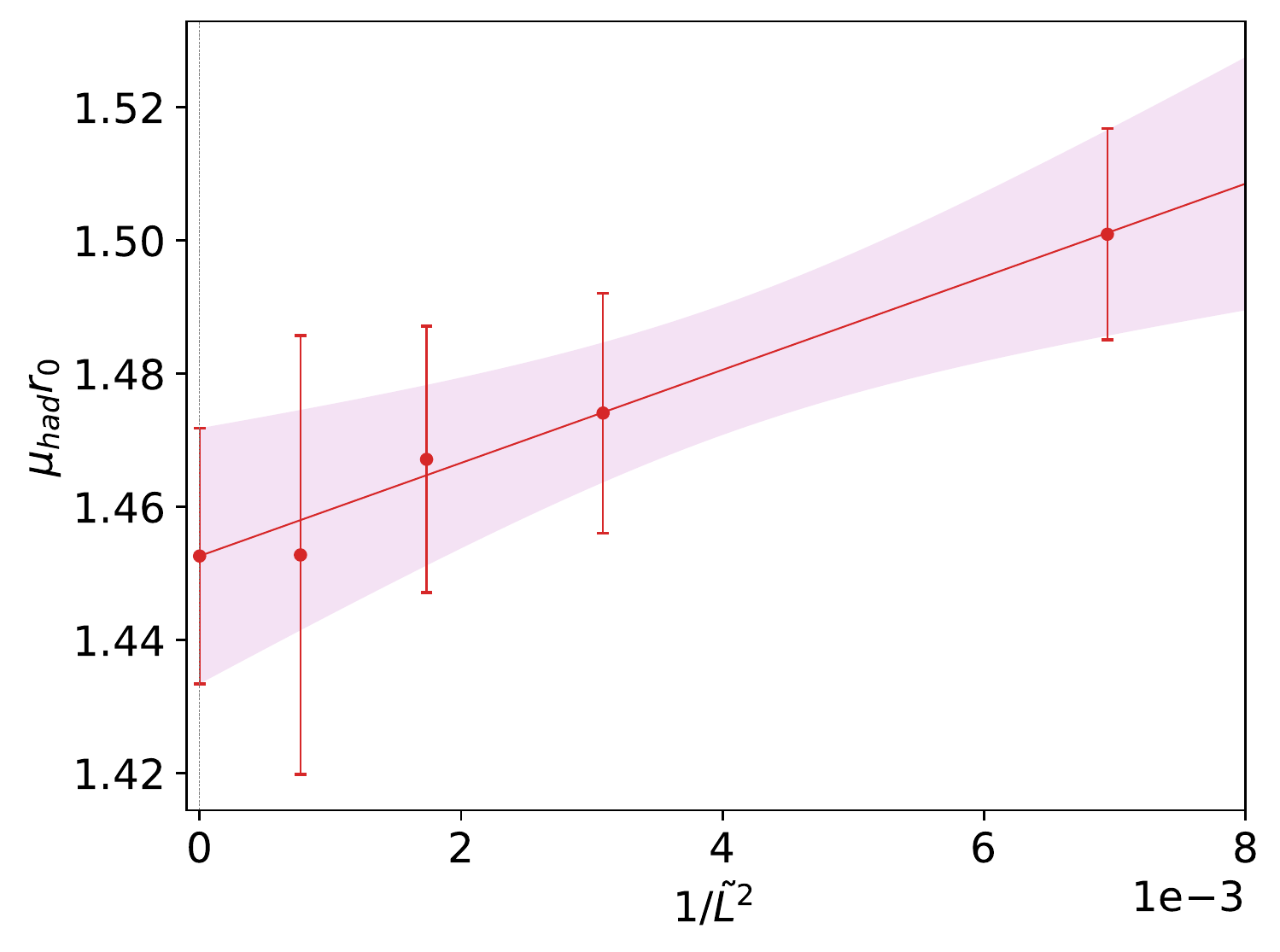}
\caption{$\mur  \times r_0$. }
\label{r0}
\end{subfigure}
\caption{Continuum extrapolation of the reference scale $\mur$ in units of $t_0$ and $r_0$.}
\end{figure}

After this, one has to determine $a/\sqrt{8 t_0}$ at those values of the bare coupling.
Following ref.~\cite{DallaBrida:2019wur}, this is done by using the results quoted in
refs.~\cite{Luscher:2010iy,Giusti:2018cmp,Knechtli:2017xgy}. All the
values for this quantity given in table 4 of ref.~\cite{DallaBrida:2019wur} are used to fit $\log(t_0/a^2)$ 
to a polynomial in $b$ of degree 5 in the range $b\in[0.33,0.39]$. This fit is then used to determine 
$a/\sqrt{8 t_0}$ at the specific values of the bare coupling  
corresponding to $\lambda(\mur)$ on each lattice. The results obtained from this procedure on lattices with
$\tL = 12$, 18, 24, 36 and 48 are given in table~\ref{t0fits}.

\begin{figure}[t] \centering
\includegraphics[width=\textwidth]{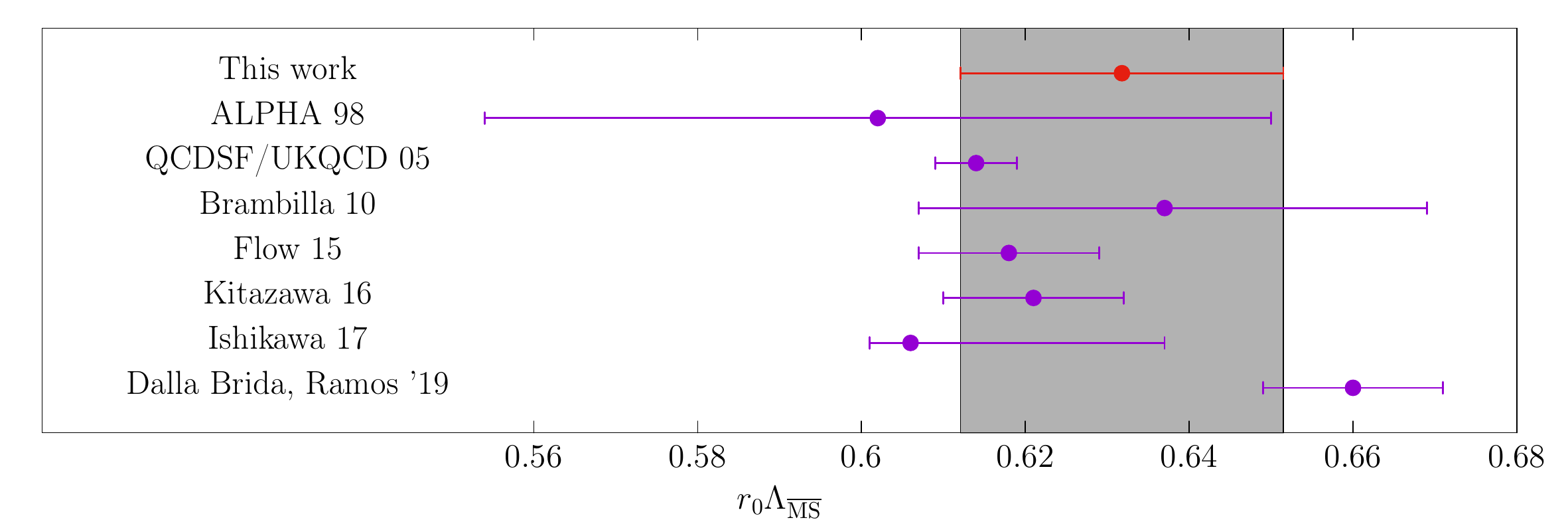}
\caption{Final result for $r_0\Lambda_{\MS}$ compared with
the values included in the FLAG average~\cite{Aoki:2019cca}. These are: ALPHA
98~\cite{Capitani:1998mq}, QCDSF/UKQCD 05~\cite{Gockeler:2005rv},
Brambilla~\cite{Brambilla:2010pp}, Flow 15~\cite{Asakawa:2015vta},
Kitazawa 16~\cite{Kitazawa:2016dsl}, Ishikawa
17~\cite{Ishikawa:2017xam}. 
We also include the recent work Dalla Brida, Ramos 19~\cite{DallaBrida:2019wur} in the comparison. 
}
\label{Figure_LMS}
\end{figure}

Figure~\ref{t0} shows the continuum extrapolation of $\mur\sqrt{8t_0}$
obtained using this strategy.  We include datapoints corresponding to
the $\tL=18$, 24, 36 and 48 lattices.  The final continuum extrapolated value is given by:
\begin{equation} 
\mur \times \sqrt{8t_0} = 1.3860(62).
\label{eq.t0result}
\end{equation}

We can now use these results to express $\Lambda_\MS$ in units of
$\sqrt{8t_0}$. Our final result (cf. eq.~\eqref{eq.Lambda_muhad}) is:
\begin{equation}
\sqrt{8t_0} \times   \Lambda_{\overline{MS}} = 0.603(17).
\label{t0eq_final}
\end{equation} 
This number is in agreement within errors with a more precise determination obtained in a recent work 
by one of the present authors~\cite{DallaBrida:2019wur}: 
$\sqrt{8t_0} \times   \Lambda_{\overline{MS}} = 0.6227(98)$. The strategy to obtain the $\Lambda$ parameter in that work is similar to the one presented here but is based on the use of a different gradient flow scheme 
with SF instead of twisted boundary conditions.

A more detailed comparison with the existing literature is better done in terms of the scale 
$r_0$~\cite{Sommer:1993ce}, which, although less precise that $t_0$, is used in most of the determinations
of the $\Lambda$ parameter up to date.
An analysis completely analogous to the one leading to eq.~\eqref{t0eq_final}, allows to 
determine $r_0 \times \Lambda_\MS$. The values of $r_0/a$ employed in
the extrapolation are shown in table~\ref{tab:t0data}. 
The values of the lattice spacing $a/r_0$ corresponding to the
renormalization condition $\lambda(\mur)$ are  
shown in table~\ref{t0fits} and the continuum extrapolation is 
displayed in fig.~\ref{r0}, leading to a value of $\mur \times r_0 = 1.453(19)$.
Our final result for the $\Lambda$ parameter in units of $r_0$ is:  
\be
r_0 \times \Lambda_\MS= 0.632(20) \, .
\ee 

Figure~\ref{Figure_LMS} shows a comparison of this result with the values included in the 
FLAG average~\cite{Aoki:2019cca}. With the present accuracy, our result comes out compatible with both the FLAG average  $r_0 \times \Lambda_\MS= 0.615(18)$, and the result by Dalla Brida and Ramos: $r_0 \times \Lambda_\MS= 0.660(11)$~\cite{DallaBrida:2019wur}. We believe that our particular scheme can be efficiently used to pin down the error in the determination of the $\Lambda$ parameter and clarify the tension among the existing values in the 
literature. 


\section{The effect of topology}
\label{sec:topology}
\begin{table}[b]
\centering
\begin{tabular}{llll}
  \toprule
$Q$ &   $\lambda(\tL=24)$ & $\lambda(\tL=36)$  & $\lambda(\tL=48)$ \\
 \midrule
0 &32.18(10)&33.29(19)  & 34.00(19)\\
1 & 42.92(25) & 43.85(56) & 43.94(49)\\
-1& 43.07(20) & 44.11(50) & -\\
2 & $59.1^*$ & $56.6^*$ & -\\
-2 &$56.4^*$ & - & - \\
\bottomrule
  \end{tabular}
  \caption{Values of the TGF coupling measured on sectors with fixed value of the topological charge $Q$.
Entries with a $*$ correspond to cases where the number of configurations with charge $Q$ is less than 20 and a correct estimation of errors is not viable.
}  \label{tab:qcorr}
\end{table}

\begin{figure}[t]
\centering
\includegraphics[width=\textwidth]{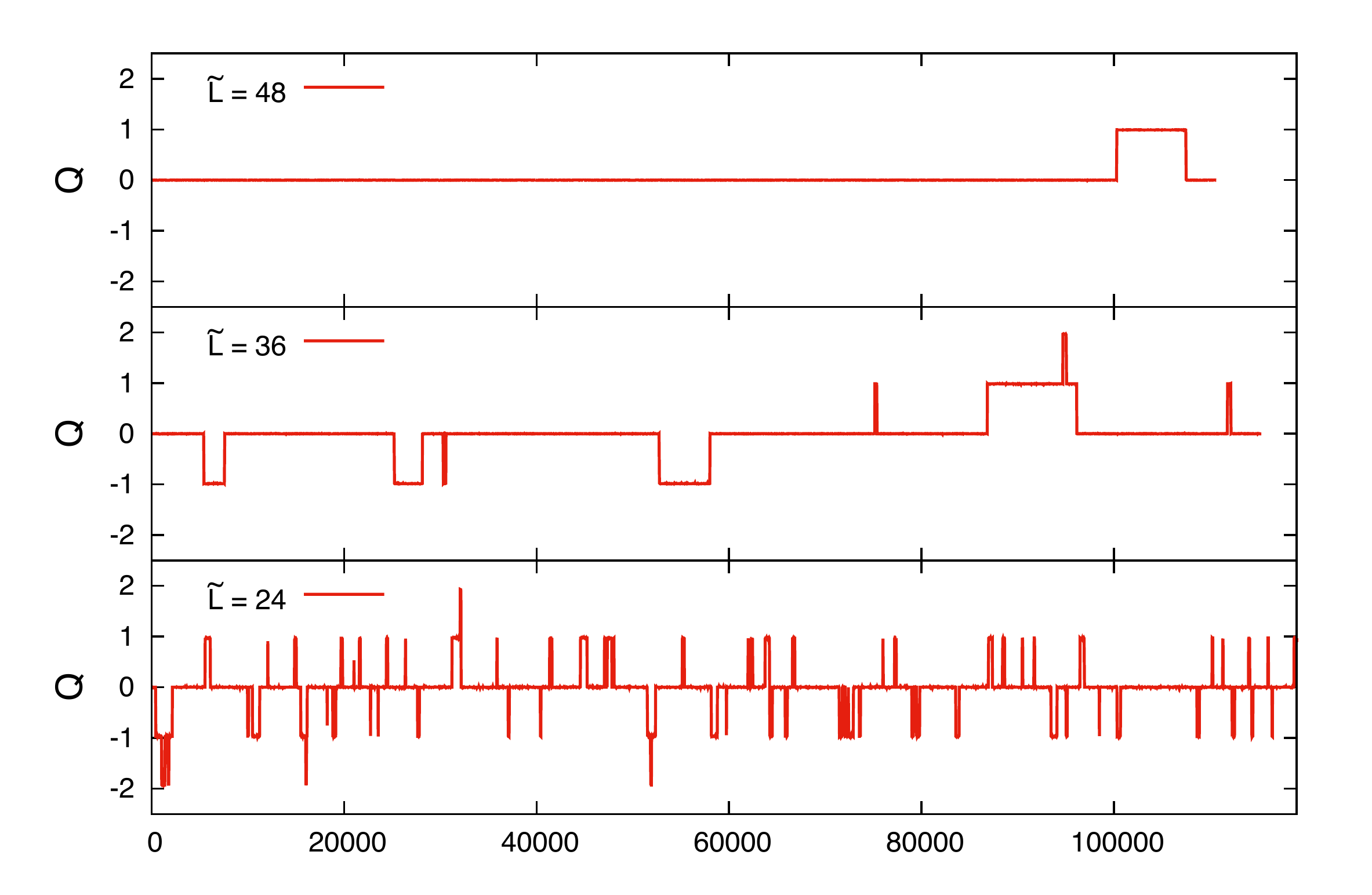}
\caption{Topological charge vs Monte Carlo time for, from top to bottom, $\tL=48$, 36 and 24, corresponding
to an approximately fixed physical size of $\tl \sim 1.1$ fm ($\lambda \sim \lambda(\mur)$) and
to lattice spacings: $a=0.023524(96)$, $0.031141(11)$, $0.045691(92)$ fm respectively.}
\label{fig:lm-top}
\end{figure}

As explained in the introduction, our definition of the gradient flow coupling follows the prescription adopted
in ~\cite{Fritzsch:2013yxa} which amounts to evaluate the coupling in the sector of trivial topology. This choice can be seen as a particular renormalization scheme,
introduced to circumvent the well known problem of toplogy freezing and critical slowing down~\cite{Schaefer:2010hu}.
A neat example of this effect is given in fig.~\ref{fig:lm-top} where we display the  value of the 
topological charge as a function of Monte Carlo time for lattices with $\tL=24,36,48$ and a physical size 
corresponding to $\lambda \approx \lambda(\mur)$ ($\tl \sim 1.1$ fm). It is clear that our statistics in the 
lattices with finer lattice spacing is not enough to correctly sample the topological charge. 
Furthermore, as table~\ref{tab:qcorr} illustrates, the correlation                                
between the coupling and the topological charge in these ensembles is very strong:
values of the coupling averaged over the $Q=0$ or $|Q|=1$ sectors differ by as much as $30\%$.
Clearly, when topology is poorly sampled, any attempt to compute the coupling in this 
volume regime by averaging over all topological sectors would lead to biased results. 

In the remaning of this section we will argue that even when topology
is frozen, topological charge fluctuations are correctly sampled in
the trivial topology 
sector (see~\cite{Albandea:2021lvl} for a study in the Schwinger
model).  We will also make an attempt to explain the observed 
dependence of the coupling on topology in semiclassical terms.  

\subsection{Topological charge fluctuations in the sector of trivial topology}

\begin{figure}[t]
\centering
\begin{subfigure}{.5\textwidth}
  \centering
\includegraphics[width=\textwidth]{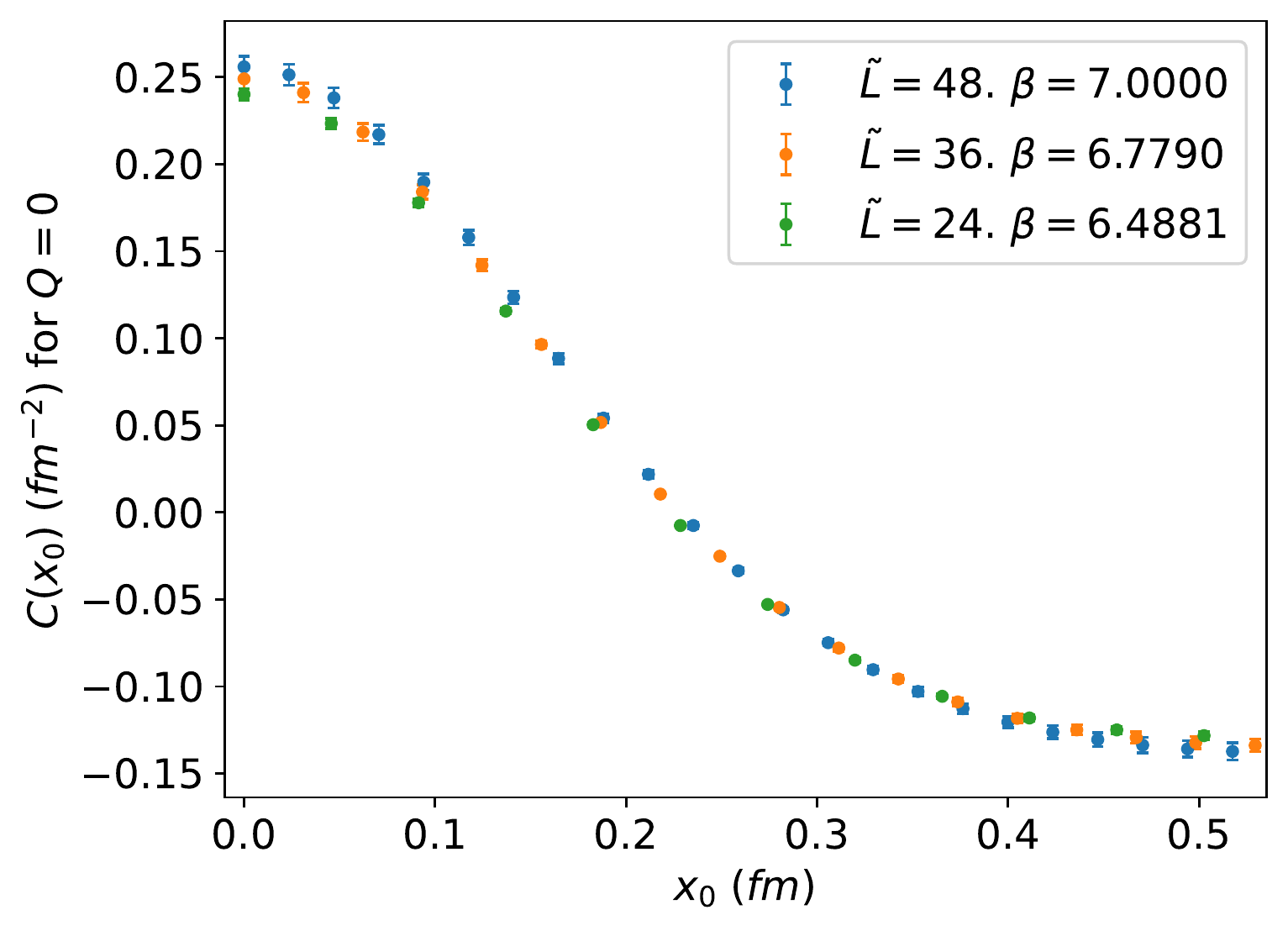}
\caption{$\tl\sim 1.1$ fm.}
\label{fig:corr-topa}
\end{subfigure}%
\begin{subfigure}{.5\textwidth}
  \centering
\includegraphics[width=\textwidth]{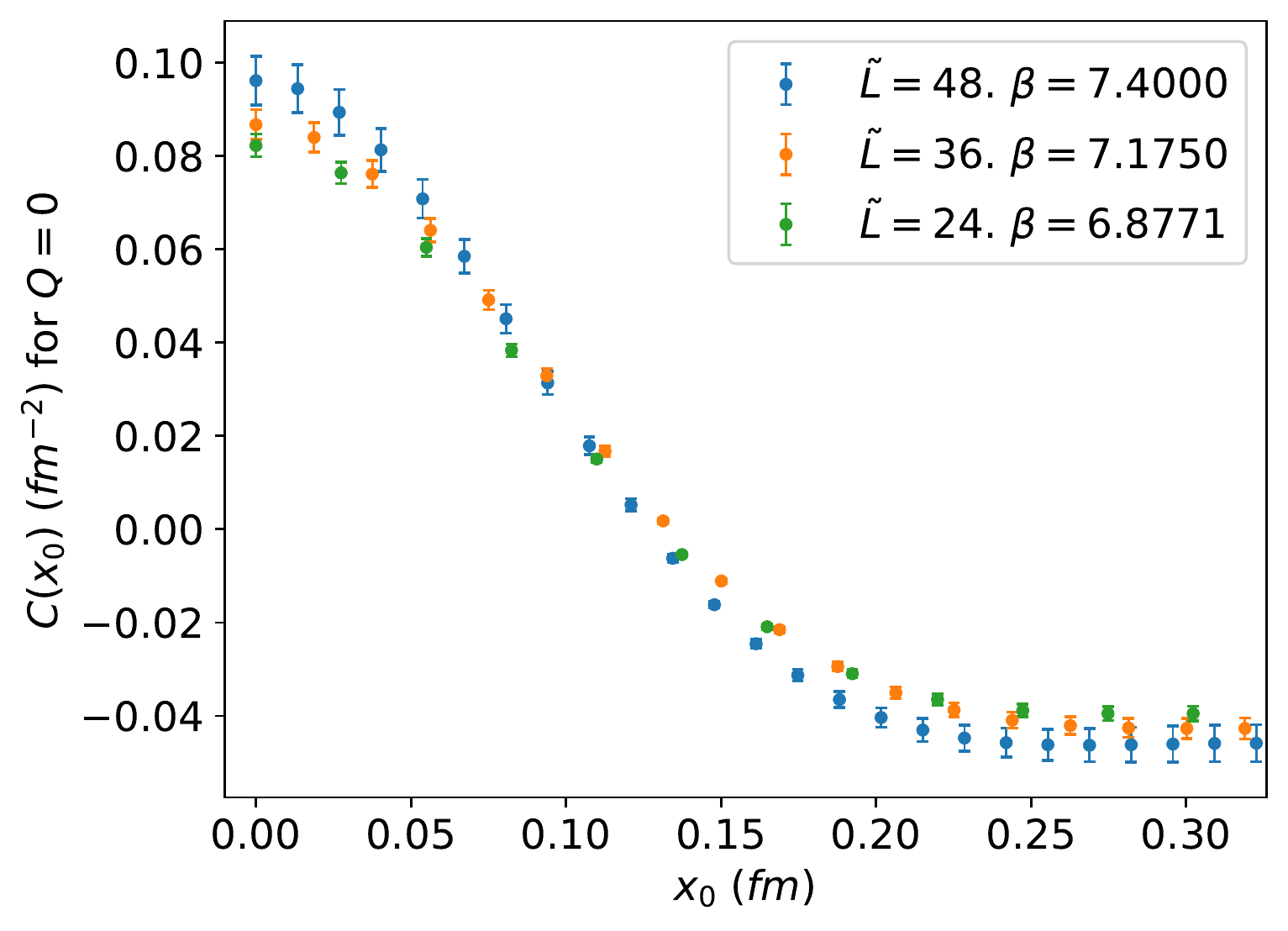}
\caption{$\tl\sim 0.66$ fm. }
\label{fig:corr-topb}
\end{subfigure}%
\caption{ We display the correlator of topological charge in the trivial topology sector, defined in eq.~\eqref{eq:qcorr}.}
\label{fig:corr-top}
\end{figure}

In order to determine whether local fluctuations of topological charge are correctly sampled in the trivial topology sector, we have computed 
the two-point function of the operator:
\be
q(x_0,t) =  \int d \vec x  \, q(\vec x, x_0, t ),
\ee
obtained by integrating the topological charge density, smeared over a flow time $t$, over three directions including the two short ones. 
On the lattice, this operator has been determined using the field theoretical definition of the charge density, cf. eq.~\eqref{eq.gmunu},
\be
q(x_0=n_0 a, t) =  \frac{a^{-1}}{16 \pi^2}  \sum_{\vec n} \Tr \left \{
G^{\rm cl}_{\mu\nu}(\vec n, n_0,t) \widetilde G^{\rm cl}_{\mu\nu}(\vec n, n_0,t) \right \} 
\, ,
\ee 
evaluated at flow time $\sqrt{8 t}= 0.3 \tl$.
In terms of this operator, we computed the correlator:
\be
C(x_0) =  \frac{\langle q(x_0, t) \, q(0,t ) \, \hat \delta_Q \rangle}{ \langle \hat \delta_Q\rangle} \Big |_{\sqrt{8 t}= 0.3 \tl}
\label{eq:qcorr}
\ee
defined over the sector with trivial topology. 

Figure~\ref{fig:corr-topa} shows the time dependence of the correlator evaluated on the Monte Carlo ensembles displayed in
fig.~\ref{fig:lm-top}. Even though topology is frozen on the largest lattice with $\tL=48$, we observe no significant deviation from scaling; the structure of the 
charge correlation is the same on the three lattices.
These good scaling properties are independent of the physical size of the box and preserved at smaller volumes, an example for  $\tl\sim 0.66$ fm is displayed in fig.~\ref{fig:corr-topb}.
These results indicate that local fluctuations of topological charge are correctly 
implemented in the trivial topology sector, even when the total topological charge is frozen to zero
in the Monte Carlo ensemble.

\subsection{Semiclassical picture}
\label{sec:semi}

The step scaling sequence interpolates between the perturbative domain, where non-trivial topological charge is suppressed, and the non-perturbative, strongly coupled regime. In between, in 
the small to intermediate volume domain, one expects semiclassical arguments based on instantons to hold. 
In this subsection those arguments will be used in an attempt to reproduce some of the features observed 
in the simulations, in particular the correlation between coupling and topological charge. 
By no means should this discussion be interpreted as a rigurous proof but just as a compelling qualitative picture of the 
role that instantons may play in this context.

In appendix~\ref{sec:diluteg} we discuss how to apply the dilute instanton gas picture to compute the leading 
semiclassical contribution to the TGF coupling in a fixed topological charge sector. 
The derivation is based on the observation that the flow removes short distance fluctuations but 
preserves classical solutions of the Euclidean equations of motion. This includes
instantons although not instanton-anti-instanton pairs, the latter will only be preserved at typical distances larger than $\sqrt{8 t}=c \tl$. 
The leading contribution to the TGF coupling in the semiclassical approximation can be estimated taking into account that:
\be
\langle E(t) \rangle =  \frac{1}{2} \langle \Tr \left (G_{\mu \nu}^2(x,t)\right) \rangle = \frac{1}{V} \left \langle \half \int d^4 x \Tr \left (G_{\mu \nu}^2(x,t)\right) \right \rangle \, .
\ee
This expression can be combined with the fact that the classical Euclidean action of a configuration with well separated instantons and anti-instantons can be approximated 
by $S_0 (n+\bar n)$, with $n(\bar n)$ the number of instantons(anti-instantons) and $S_0$ the one-instanton action.
Therefore, at lowest order in the dilute gas approximation, the energy density receives a 
(semi-)classical contribution that is
proportional to the average number of instantons plus anti-instantons, i.e.,   
\be
\langle E(t) \rangle^{\rm dig} \sim \frac{S_0}{V} \, \langle n +\bar n \rangle + \cdots \, .
\ee
This contribution adds to the perturbative one.
One can as well evaluate in this approximation the average value of any observable, 
in particular $\langle n +\bar n \rangle$, restricted to the sector of topological charge $\tilde Q$ by using:
\be
\langle O \rangle_{\widetilde Q} = \frac{\langle O\,  \delta_{Q-\widetilde Q}
\rangle }{\langle \delta_{Q-\widetilde Q} \rangle},
\label{eq:obsQ}
\ee
with $Q$ the total topological charge, in analogy to what we did for the definition of the coupling in the 
zero topological sector in eq.~\eqref{eq:lambdat}.

In order to determine the average number of instantons in the dilute gas approximation, we have considered two 
possibilities: a dilute gas of ordinary instantons of action $S_0 = 8\pi^2$, and one made out of fractional 
instantons~\cite{tHooft:1981nnx} with instanton action $S_0 = 8\pi^2/N$.  
The derivation follows the standard treatment, see i.e.
\cite{cl:symmetry}. In this context, the semiclassical picture based on the less standard case of fractional 
instantons follows the ideas put forward by Gonz\'alez-Arroyo and collaborators in a series of 
works~\cite{GarciaPerez:1993ab,GarciaPerez:1993jw,GonzalezArroyo:1995ex,GonzalezArroyo:1995zy,Perez:2010jx}, 
see also~\cite{vanBaal:1984ra,vanBaal:2000zc,Unsal:2020yeh}.

Details of the calculation are provided in appendix ~\ref{sec:diluteg}, here we will 
only summarize the main final formulas.
The contribution to the coupling derived in this approximation, when restricted to the sector of topological 
charge $Q$, reads:
\be
\lambda_{Q}^{\rm dig} (c) =  \frac{ 2 N S_0 }{ 3 \widehat {\cal A}(\pi c^2)}  \, \left (\nu + x \,  \frac{I_{\nu+1}(x)}{I_{\nu}(x)}
\right) + \cdots \, ,
\label{eq:lambda_sm}
\ee
with $\nu=Q$ for ordinary instantons, and $\nu=NQ$ for fractional ones, and where $\widehat {\cal A}(x)={\cal A}(x) /x^2$, coming from the coupling normalization, 
cf. eq.~\eqref{eq:lambdat}.
In this expression, $I_\nu(x)$ stands for the modified Bessel function of the first kind, and $x=2RV$, with $R$ the probability 
to generate one instanton, or anti-instanton, per unit volume.

\begin{figure}[t]
\centering
\includegraphics[width=0.9\textwidth]{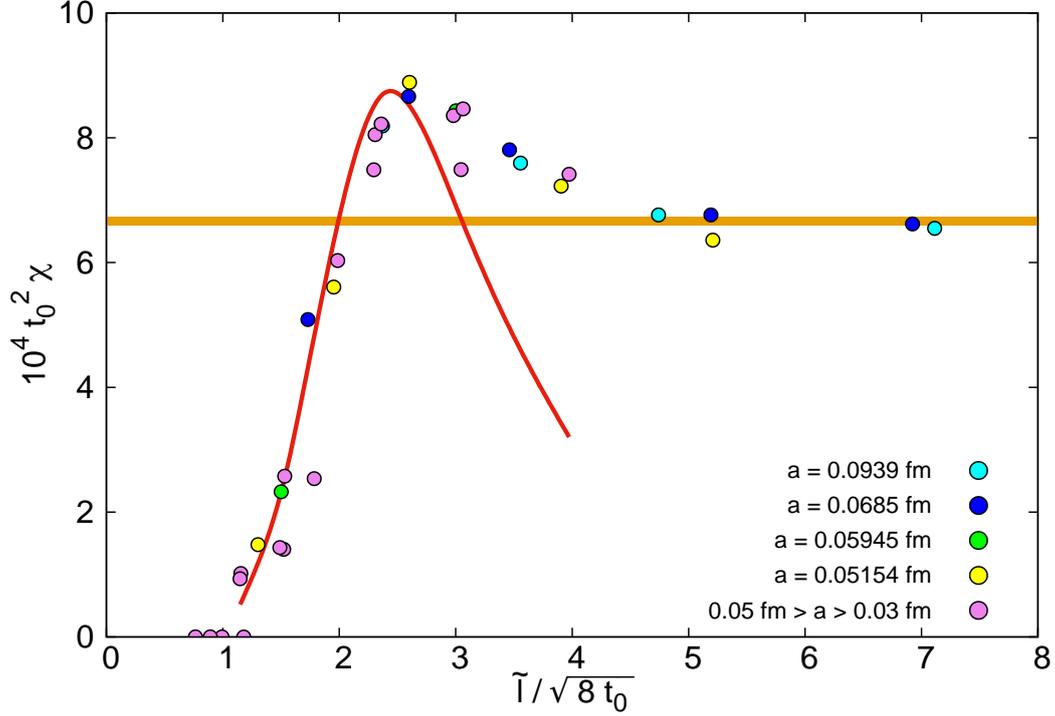}
\caption{Volume dependence of the topological susceptibility evaluated from flowed fields at 
$8 t = (0.3 \tl)^2$. The orange horizontal band corresponds to the value of the, continuum extrapolated, large-volume $SU(3)$ susceptibility
$t_0^2 \chi= 6.67(7) \times 10^{-4}$~\cite{Ce:2015qha}.
The red line gives the dilute gas estimate discussed
in sec.~\ref{sec:semi}, cf. eq.~\eqref{eq:xdiluteg} with $C = 880$. 
}
\label{fig:vol-dep-chi}
\end{figure}

\begin{figure}[t]
\centering
  \centering
\includegraphics[width=0.9\textwidth]{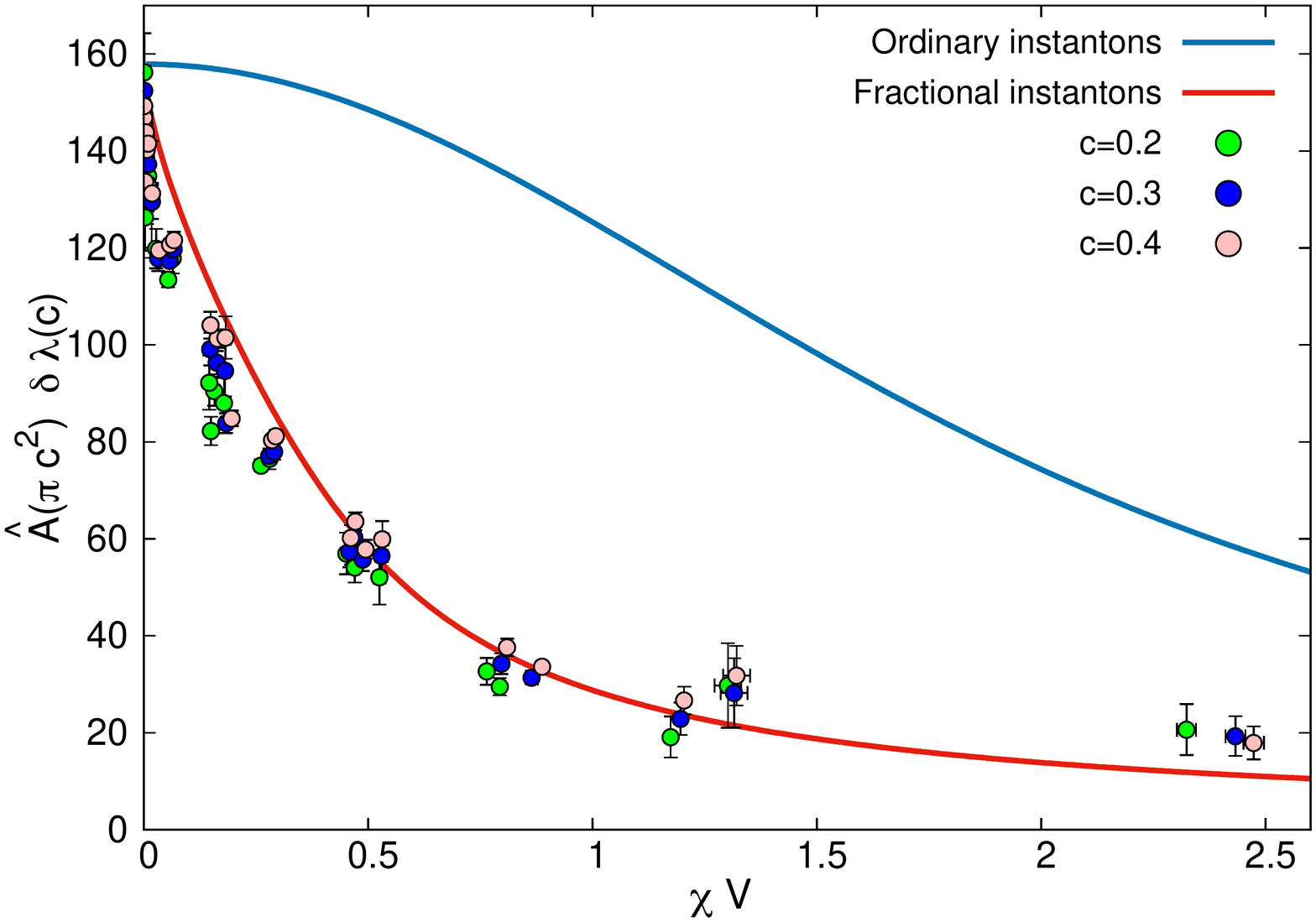}
\caption{ We display  $\widehat {\cal A}(\pi c^2)  \delta \lambda(c)= \widehat {\cal A}(\pi c^2) 
\left (\lambda_{Q=1}(c) - \lambda_{Q=0}(c)\right)$
as a funcion of $\chi V$ for couplings obtained at flow time $8t=(c\tl)^2$, with $c=0,2$, 0.3 and 0.4. The
lines correspond to the estimate using the leading semi-classical contribution from
either fractional or ordinary instantons. Note that these lines are not fits to the data, they are directly given by eq.~\eqref{eq:lambda_sm} with the parameter $x$ determined from the measured value of the susceptibility.
}
\label{fig:deltal}
\end{figure}

A way to determine $R$ from the numerical simulations is to measure the topological susceptibility.
In the dilute gas approximation and for the case of ordinary instantons, the relation between $R$ and $\chi$ is
simply given by: $\chi=2R$, while for fractional $SU(3)$ instantons one obtains instead:
\be
\chi   =  \frac{2R}{9} \left (1 - \frac{3(1+ R V )}{2+ e^{3 R V }}\right)\, .
\label{eq:chifrac}
\ee
The topological susceptibility is defined formally in the continuum through the correlator of the topological charge density $q(x)$:
\begin{equation}
       \chi=\int d^4x\left< q(x)q(0)\right>.
\end{equation}
A proper definition requires to regulate the short-distance singularity in this expression, and this can be attained by evaluating
the charge density at non-zero flow time.
On the lattice, in the absence of topology freezing, the topological susceptibility can be computed by measuring:
\be
         \chi= \frac{ \langle Q^2 \rangle }{V}
\ee
with $Q$ the total topological charge, determined from the field theoretical definition eq.~\eqref{eq:qlatt}, evaluated on the flowed
gauge fields. 

An illustration of the volume dependence of the topological susceptibility is provided
in fig.~\ref{fig:vol-dep-chi} where we display $t_0^2\chi$ as a function of $\tl /\sqrt{8 t_0}$. 
The results shown correspond to simulations with lattice spacing $a > 0.03$ fm (we refrain to give errors 
in this plot since our determination of the susceptibility may be affected by freezing, in particular 
for values of $a<0.05$ fm).
The onset of the susceptibility is very abrupt and takes place for values of $\tl/ \sqrt{8 t_0} \in [1, 3]$,
overlapping with the region covered by the step scaling sequence.
After that, $\chi$ stabilizes to the constant large volume value, given in $SU(3)$ by $t_0^2 \chi= 6.67(7) \times 10^{-4}$~\cite{Ce:2015qha}, corresponding in the figure to the orange horizontal band.

By fixing $x$ in terms of the measured susceptibility, one can 
evaluate the dilute gas contribution to $\delta \lambda(c)= \lambda_{Q=1}(c) - \lambda_{Q=0}(c)$ and compare 
the result to the values of $\delta \lambda(c)$ determined in the simulations. 
This has been done for the set of simulations with $\beta \in [6,6.8]$
where topology seems to be properly sampled and the number of $Q=1$ configurations is sufficiently large.
In addition, the calculation has been done for various values of the flow time: $8 t = (c \tl)^2$, with
$c=0.2$, 0.3 and 0.4.
We observe that the quantity $\widehat {\cal A}(\pi c^2) \delta \lambda(c)$
is quite independent of the value of $c$. 
Our results are displayed
as a function of $\chi V$ in fig.~\ref{fig:deltal} and compared with the dilute gas predictions 
for ordinary and fractional instantons, derived from eq.~\eqref{eq:lambda_sm}.
Surprisingly, the dependence predicted by the fractional instanton formula 
matches remarkably well our results in a large range of volumes. 
We stress that the lines displayed in the figure do not have any 
free parameter, they are directly given by eq.~\eqref{eq:lambda_sm} with the input parameter $x$ fixed 
in terms of the measured value of the susceptibility. 

In the dilute gas approximation one can as well obtain a prediction for the parameter $x=2RV$ in terms 
of the renormalized coupling. 
As explained in appendix~\ref{sec:diluteg}, the result for fractional instantons at one loop order reads:
\be
x=2RV =  C \, \frac{V}{\rho^4}\, \left(\lambda(1/\rho)\right)^{-2} \,  e^{- 8\pi^2 / \lambda(1/\rho)}
\label{eq:xdiluteg}
\ee
where $\rho$ stands for the characteristic fractional instanton size, 
proportional to $\tl$ in the small volume regime, 
and $C$ for a normalization constant, unfortunately unknown. 
Disregarding this fact, we have used this functional form to get a qualitative idea of 
the volume dependence of the topological susceptibility in this approximation.
The calculation requires to input the renormalized coupling constant at a scale $\rho \propto \tl$; differences
in the scheme dependence translate at this order in a change of value of $C$ that depends
on the ratio of $\Lambda$ parameters. We have tested the formula setting $\rho= c \tl$ and using the coupling
in the TGF scheme with $c=0.2$, 0.3 and 0.4. The one with $c=0.2$ is describing better
our small volume results; the red line displayed in fig.~\ref{fig:vol-dep-chi} corresponds to this
value of $c$ and $C = 880$.

To summarize, although the results presented in this section have to be taken with a grain of salt, as they are not based on a rigurous analytic calculation, we believe they provide compelling indication that some of the observed effects, in
particular the correlation between topological charge and coupling, can be understood in semiclassical terms.


\section{Conclusions}
\label{sec:conclusions}
In this work we have investigated the running coupling using finite
size scaling techniques in a scheme with twisted boundary conditions. 
We have used a particular choice of geometry, motivated by the study
of the large $N$ limit of $SU(N)$ Yang Mills theories. 
In this scheme the physical size of the four dimensional volume
is asymmetric, with two directions being a factor $N$ larger than the
other two. 

We have focused our investigations in the group $SU(3)$.
For this case there are many results available in the literature,
which allows for a more stringent comparison. Our determination of the
$\Lambda_{\overline{\rm MS} }$
parameter in units of the flow scale $\sqrt{8t_0}$ or the Sommer scale $r_0$ shows a good
agreement with the literature.
The precision of our computation does not match the one of the most precise determinations,
but the systematic associated with the continuum
extrapolation is taken into account seriously and our simulations involve large lattices.
In addition, the matching with perturbation theory is performed in several ways,
using a non-perturbative running that covers large energy scales.

In this respect, it is important to note that the determination of $\Lambda_{\overline{\rm MS} }$
is accomplished in two different ways. First, we perform a direct determination
in our finite volume scheme with twisted boundary conditions.
In this case the ratio $\Lambda_{\overline{\rm MS} }/\Lambda_{\rm
  TGF}$ is a key piece in the computation.
Second, we perform a non-perturbative matching with the SF coupling,
using available data from the literature.
The two determinations show a good agreement, but they exhibit a subtle point.
The extraction of the $\Lambda$ parameter at a certain energy scale
$\mupt$ has perturbative corrections $\mathcal O(\lambda^k(\mupt))$
(with $k=1$ in the case of the direct extraction, and
$k=2$ for the case of the non-perturbative matching with the SF).
A crucial point to obtain compatible results is to remove these subleading
perturbative corrections by performing an extrapolation $\lambda(\mupt) \to 0$.
At our level of precision, and specially for the case of the
direct extractions, perturbative corrections are significant, even at
energy scales ridiculously large.

All in all, this work shows two promising directions for future research. 
First, extracting the $N$ dependence of the $\Lambda$-parameter in
this particular scheme is computationally cheaper than a brute force approach. 
Second, the particular choice of geometry shows a smaller memory
footprint than usual simulations with symmetrical lattices. This
smaller memory footprint implies an increase in the ratio computing
over memory transfer, and therefore can be executed more efficiently
in current hardware accelerators. Here it is important to stress
that this pure gauge case can be non-perturbatively matched to QCD using heavy quarks~\cite{DallaBrida:2019mqg}
and has therefore direct phenomenological relevance for the extraction of the strong coupling in QCD.

The authors hope to investigate these points in detail in future works.


\section*{Acknowledgments}
\addcontentsline{toc}{section}{Acknowledgments}

We would like to thank F.~Arco, J.L.F.~Barb\'on, J.~Calder\'on~Infante, A.~Gonz\'alez-Arroyo
and C.~Pena for many useful discussions
on this and related topics.  
AR acknowledges financial support from the Generalitat Valenciana
(genT program CIDEGENT/2019/040). EIB, JDG and MGP acknowledge
financial support from the MINECO/FEDER grant FPA2015-68541-P and
PGC2018-094857-B-I00 and the MINECO Centro de Excelencia Severo Ochoa
Program SEV-2016-0597. E.I. Bribi\'an acknowledges support under the
FPI grant BES-2015-071791. J. Dasilva Gol\'an acknowledges support under the FPI grant
PRE2018-084489.  This publication is supported by the European project
H2020-MSCAITN-2018-813942 (EuroPLEx) and the EU Horizon 2020 research
and innovation programme, STRONG-2020 project, under grant agreement
No 824093.  The numerical computations have been carried out on the
Hydra cluster at IFT and with computer resources provided by CESGA
(Supercomputing Centre of Galicia).

\cleardoublepage

\appendix

\section{Lattice ensembles}
\label{sec:esembles}
\begin{longtable}[hp]{c | c | c | c | c | c}
\toprule
$18\times b$ & $\tilde{L}=12$ & $\tilde{L}=18$ & $\tilde{L}=24$ & $\tilde{L}=36$ & $\tilde{L}=48$\\
\midrule
6.0000	&	7517(2483)	&20248(30694)	&	3920(13062)	&1070(8702)	&	120(1876)\\
6.2000	&	9569(431)	&36970(15606)	&	7869(10087)	&1120(4159)	&	220(1713)\\
6.2980	&	9887(113)	&	&	5644(4356)	&	&	\\
6.4000	&	9961(39)	&49359(3417)	&	13050(5356)	&	3899(6983)	&	458(1541)\\
6.4881	&	9983(17)	&	&	8350(1650)	&	&	\\
6.5837	&	&	9940(60)	&	&	3556(2311)	&	\\
6.6000	&	10000(0)	&52807(300)	&	16713(1287)	&6476(4482)	&759(1537)	\\
6.6816	&	10000(0)	&	&	9805(195)	&	&	\\
6.7790	&	&	10000(0)	&	&	2634(569)	&	\\
6.8000	&	10000(0)	&51840(67)	&	17807(193)	&9593(1616)	&1316(989)	\\
6.8771	&	10000(0)	&	&	10000(0)	&	&	\\
6.9769	&	&	10000(0)	&	&	4344(1615)	&	\\
7.0000	&	10000(0)	&53238(0)	&	16439(0)	&10971(0)	&2155(150)	\\
7.0739	&	10000(0)	&	&	10000(0)	&	&	\\
7.1750	&	&	10000(0)	&	&	3198(0)	&	\\
7.2000	&	10000(0)	&50544(0)	&	18000(0)	&10974(0)	&2307(0)	\\
7.2713	&	10000(0)	&	&	10000(0)	&	&	\\
7.3743	&	&	10000(0)	&	&	3231(0)	&	\\
7.4000	&	10000(0)	&53406(0)	&	16763(0)	&10981(0)	&2603(0)	\\
7.4691	&	10000(0)	&	&	10000(0)	&	&	\\
7.5735	&	&	10000(0)	&	&	3219(0)	&	\\
7.6000	&	10000(0)	&	53605(0)	&	17760(0)	&9999(0)	&	2308(0)	\\
7.7734	&	&	10000(0)	&	&	3240(0)	&	\\
7.8000	&	10000(0)	&53560(0)	&	17501(0)	&8979(0)	&2604(0)	\\
7.9663	&	10000(0)	&	&	10000(0)	&	&	\\
8.0000	&	10000(0)	&53602(0)	&	18000(0)	&9720(0)	&	2610(0)\\
8.2721	&	&	10000(0)	&	&	3229(0)	&	\\
8.4643	&	10000(0)	&	&	10000(0)	&	&	\\
8.5000	&	10000(0)	&53825(0)	&	17000(0)	&8990(0)	&2324(0)	\\
8.7711	&	&	10000(0)	&	&	3237(0)	&	\\
8.9629	&	10000(0)	&	&	10000(0)	&	&	\\
9.0000	&	10000(0)	&53654(0)	&	18000(0)	&9000(0)	&2333(0)	\\
9.2704	&	&	10000(0)	&	&	1602(0)	&	\\
9.4615	&	10000(0)	&	&	10000(0)	&	&	\\
9.5000	&	10000(0)	&54178(0)	&	16483(0)	&7998(0)	&1966(0)\\
9.7694	&	&	10000(0)	&	&	3255(0)	&	\\
9.9609	&	10000(0)	&	&	10000(0)	&	&	\\
10.0000	&	10000(0)	&53889(0)	&	18000(0)	&13866(0)	&	2898(0)	\\
10.2686	&	&	10000(0)	&	&	8159(0)	&	\\
10.4612	&	10000(0)	&	&	10000(0)	&	&	\\
10.5000	&	10000(0)	&53415(0)	&	17000(0)	&	12841(0)	&2898(0)\\
10.7678	&	&	10000(0)	&	&	7714(0)	&	\\
11.0000	&	10000(0)	&53679(0)	&	8640(0)	&12800(0)	&2584(0)	\\
11.4607	&	10000(0)	&	&	10000(0)	&	&	\\
11.7682	&	&	10000(0)	&	&	6547(0)	&	\\
12.0000	&	10000(0)	&10000(0)	&	16442(0)	&8916(0)	&1658(0)	\\
12.4606	&	10000(0)	&	&	10000(0)	&	&	\\
12.7677	&	&	10000(0)	&	&	6550(0)	&	\\
13.0000	&	10000(0)	&10000(0)	&	16513(0)	&8873(0)	&1655(0)	\\
13.4606	&	10000(0)	&	&	10000(0)	&	&	\\
13.7673	&	&	10000(0)	&	&	5643(0)	&	\\
14.0000	&	10000(0)	&10000(0)	&	16340(225)	&8400(0)	&1663(0)	\\
14.4612	&	10000(0)	&	&	10000(0)	&	&	\\
14.7634	&	&	10000(0)	&	&	8158(0)	&	\\
15.0000	&	10000(0)	&10000(0)	&	16602(0)	&8247(0)	&1668(0)	\\
20.0000 &	10000(0)	&5009(0)		&	11570(0)&1893(0)&410(0)\\
\bottomrule
\caption{Total number of configurations with zero topological charge
(within parenthesis those with non-zero charge).}
\label{tableA.1}
\end{longtable}

\section{Raw data}
\label{sec:raw-data}
\begin{longtable}{ c | c | c | c | c | c }
\toprule
 $18	\times b$ & $\tilde{L}=12$ & $\tilde{L}=18$ & $\tilde{L}=24$ & $\tilde{L}=36$ & $\tilde{L}=48$\\
 \midrule
6.00000	&	34.80(11)	&	70.849(96)	&	128.69(32)	&	337.0(11)	&	656.0(51)	\\
6.20000	&	21.567(67)	&	39.414(52)	&	66.69(15)	&	155.25(66)	&	310.0(22)	\\
6.29800	&	17.947(52)	&				&	50.75(15)	&				&				\\
6.40000	&	15.483(42)	&	24.809(35)	&	39.530(88)	&	84.46(24)	&	155.07(95)	\\
6.48810	&	13.947(35)	&				&	32.18(10)	&				&				\\
6.58370	&				&	17.892(50)	&				&	52.00(20)	&				\\
6.60000	&	12.448(28)	&	17.484(22)	&	25.319(60)	&	50.06(14)	&	86.88(55)	\\
6.68160	&	11.577(23)	&				&	21.746(70)	&				&				\\
6.77900	&				&	13.931(34)	&				&	33.29(19)	&				\\
6.80000	&	10.574(20)	&	13.585(14)	&	17.907(41)	&	31.836(90)	&	52.44(31)	\\
6.87710	&	9.994(18)	&				&	16.148(42)	&				&				\\
6.97690	&				&	11.573(23)	&				&	22.67(11)	&				\\
7.00000	&	9.190(15)	&	11.360(10)	&	13.916(27)	&	21.603(63)	&	34.00(19)	\\
7.07390	&	8.832(15)	&				&	12.975(30)	&				&				\\
7.17500	&				&	10.002(18)	&				&	16.596(61)	&				\\
7.20000	&	8.243(13)	&	9.8366(80)	&	11.571(17)	&	16.023(43)	&	22.78(16)	\\
7.27130	&	7.943(12)	&				&	10.933(21)	&				&				\\
7.37430	&				&	8.826(15)	&				&	13.210(43)	&				\\
7.40000	&	7.444(11)	&	8.7118(62)	&	9.999(14)	&	12.878(27)	&	16.599(89)	\\
7.46910	&	7.232(11)	&				&	9.570(17)	&				&				\\
7.57350	&				&	7.927(12)	&				&	11.170(33)	&				\\
7.60000	&	6.836(11)	&	7.8398(53)	&	8.846(11)	&	10.931(21)	&	13.370(68)	\\
7.77340	&				&	7.220(11)	&				&	9.736(24)	&				\\
7.80000	&	6.2882(91)	&	7.1400(47)	&	7.9290(95)	&	9.530(17)	&	11.229(44)	\\
7.96630	&	5.9170(84)	&				&	7.337(11)	&				&				\\
8.00000	&	5.8619(84)	&	6.5569(42)	&	7.2322(83)	&	8.482(14)	&	9.731(33)	\\
8.27210	&				&	5.9161(85)	&				&	7.452(20)	&				\\
8.46430	&	5.0260(68)	&				&	5.9971(85)	&				&				\\
8.50000	&	4.9827(67)	&	5.4767(33)	&	5.9142(66)	&	6.717(11)	&	7.467(25)	\\
8.77110	&				&	5.0356(69)	&				&	6.072(15)	&				\\
8.96290	&	4.3739(59)	&				&	5.0883(70)	&				&				\\
9.00000	&	4.3426(57)	&	4.7096(28)	&	5.0237(51)	&	5.5922(83)	&	6.092(18)	\\
9.27040	&				&	4.3850(59)	&				&	5.128(17)	&				\\
9.46150	&	3.8881(51)	&				&	4.4249(59)	&				&				\\
9.50000	&	3.8529(50)	&	4.1335(23)	&	4.3831(46)	&	4.8058(74)	&	5.146(16)	\\
9.76940	&				&	3.8813(51)	&				&	4.466(10)	&				\\
9.96090	&	3.4945(44)	&				&	3.9068(51)	&				&				\\
10.0000	&	3.4676(44)	&	3.6895(20)	&	3.8903(38)	&	4.2029(47)	&	4.480(11)	\\
10.2686	&				&	3.4943(45)	&				&	3.9500(57)	&				\\
10.4612	&	3.1746(40)	&				&	3.5160(44)	&				&				\\
10.5000	&	3.1481(38)	&	3.3331(18)	&	3.4938(34)	&	3.7424(43)	&	3.954(11)	\\
10.7678	&				&	3.1747(39)	&				&	3.5280(52)	&				\\
11.0000	&	2.8860(36)	&	3.0458(17)	&	3.1720(42)	&	3.3810(40)	&	3.5505(94)	\\
11.4607	&	2.6834(32)	&				&	2.9271(36)	&				&				\\
11.7682	&				&	2.6817(32)	&				&	2.9499(45)	&				\\
12.0000	&	2.4778(30)	&	2.5955(32)	&	2.6819(26)	&	2.8259(36)	&	2.9393(94)	\\
12.4606	&	2.3270(28)	&				&	2.5031(31)	&				&				\\
12.7677	&				&	2.3320(28)	&				&	2.5211(38)	&				\\
13.0000	&	2.1738(25)	&	2.2608(27)	&	2.3259(22)	&	2.4370(31)	&	2.5134(82)	\\
13.4606	&	2.0568(24)	&				&	2.1967(26)	&				&				\\
13.7673	&				&	2.0601(25)	&				&	2.2038(36)	&				\\
14.0000	&	1.9357(23)	&	2.0069(23)	&	2.0584(19)	&	2.1432(28)	&	2.2130(65)	\\
14.4612	&	1.8425(21)	&				&	1.9503(23)	&				&				\\
14.7634	&				&	1.8449(22)	&				&	1.9585(27)	&				\\
15.0000	&	1.7483(20)	&	1.7971(21)	&	1.8439(17)	&	1.9082(25)	&	1.9593(57)\\
20.0000	&	1.1748(13)	&	1.1988(20)	&	1.2137(13)	&	1.2440(32)	&	1.2705(68)\\
\bottomrule
\caption{TGF 't Hooft coupling data as a function of the inverse bare coupling $b=1/\lambda_0$ and the 
lattice size $\tilde{L}= \tl/a$.}
\label{tableB.1}
\end{longtable}

\section{Continuum extrapolation for the step scaling function}
\label{sec:extrp}
\label{ap.extrap}
\begin{longtable}{ c   c  c  c   c  } 
\toprule
$\tilde{L}$ & $18\times b$ & $u$ & $\Sigma(u)$ & $\Sigma(u_{\rm tg})$ \\	
\midrule
12	&	14.4612	&	1.8425(21)	&	1.9503(23) 	&	1.9503(34)		\\
18	&	14.7634	&	1.8449(22)	&	1.9566(40)  	&	1.9558(37)		\\
24	&	15.0000	&	1.8439(17)	&	1.9593(58)  	&	1.9577(61)		\\
	&		&			&			&				\\
$\infty$   	&	  	&	$u_{\rm tg}=$1.8425	&		&	 1.9602(55)  	\\
\midrule
12	&	13.4606	&	2.0568(23)	&	2.1967(26)   	&	2.1967(38)		\\
18	&	13.7630	&	2.0601(25)	&	2.2038(35)    	&	2.2002(46)		\\
24	&	14.0000	&	2.0584(19)	&	2.2130(64)    	&	2.2113(68)		\\
	&		&			&			&				\\
$\infty$   	&	  	&	$u_{\rm tg}=$2.0568	&		&	 2.2092(65)  	\\
\midrule
12	&	12.4606	&	2.3270(28)	&	2.5031(31)   	&	2.5031(45)		\\
18	&	12.7677	&	2.3320(28)	&	2.5211(38)     	&	2.5153(50)		\\
24	&	13.0000	&	2.3259(22)	&	2.5134(81)    	&	2.5147(85)		\\
	&		&			&			&				\\
$\infty$   	&	  	&	$u_{\rm tg}=$2.3270	&		&	 2.5224(75) 	\\
\midrule
12	&	11.4607	&	2.6834(32)	&	2.9271(36)   	&	2.9271(52)		\\
18	&	11.7682	&	2.6817(32)	&	2.9498(44)    	&	2.9519(59)		\\
24	&	12.0000	&	2.6819(26)	&	2.9393(93)    	&	2.9411(99)		\\
	&		&			&			&				\\
$\infty$   	&	  	&	$u_{\rm tg}=$2.6834	&		&	 2.9607(88) 	\\
\midrule
12	&	10.4612	&	3.1746(40)	&	3.5160(44)   	&	3.5160(66)		\\
18	&	10.7678	&	3.1747(39)	&	3.5280(52)    	&	3.5278(71)		\\
24	&	11.0000	&	3.1720(42)	&	3.5505(94)   	&	3.554(11)		\\
	&		&			&			&				\\
$\infty$   	&	  	&	$u_{\rm tg}=$3.1746	&		&	 3.555(10)  	\\
\midrule
12	&	9.96090	&	3.4945(44)	&	3.9068(51)   	&	3.9068(75)		\\
18	&	10.2686	&	3.4943(45)	&	3.9500(57)    	&	3.9502(81)		\\
24	&	10.5000	&	3.4938(34)	&	3.954(11)  	&	3.955(11)	\\
	&		&			&			&				\\
$\infty$   	&	  	&	$u_{\rm tg}=$3.4945	&		&	 3.978(11)	\\
\midrule
12	&	9.46150	&	3.8881(50)	&	4.4249(59)   	&	4.4249(88)		\\
18	&	9.76940	&	3.8813(51)	&	4.466(10)  	 	&4.475(12)	\\
24	&	10.0000	&	3.8903(38)	&	4.480(11)    	&	4.477(12)		\\
	&		&			&			&				\\
$\infty$   	&	  	&	$u_{\rm tg}=$3.8881	&		&	 4.501(14) 	\\
\midrule
12	&	8.96290	&	4.3739(59)	&	5.0883(70)    	&	5.088(11)		\\
18	&	9.27040	&	4.3850(59)	&	5.128(17)     	&	5.113(19)		\\
24	&	9.50000	&	4.3831(46)	&	5.146(16)      	&	5.133(17)		\\
	&		&			&			&				\\
$\infty$   	&	  	&	$u_{\rm tg}=$4.3739	&		&	 5.143(20)  	\\
\midrule
12	&	8.46430	&	5.0260(68)	&	5.9971(85)      &	5.997(13)		\\
18	&	8.77110	&	5.0356(69)	&	6.072(15)     	&	6.058(18)		\\
24	&	9.00000	&	5.0237(51)	&	6.092(18)     	&	6.095(20)		\\
	&		&			&			&				\\
$\infty$   	&	  	&	$u_{\rm tg}=$5.0260	&		&	 6.120(22)  	\\
\midrule
12	&	7.96630	&	5.9170(84)	&	7.337(11)    	&	7.337(17)		\\
18	&	8.27210	&	5.9161(85)	&	7.452(20)     	&	7.453(24)		\\
24	&	8.50000	&	5.9142(66)	&	7.467(25)    	&	7.471(27)		\\
	&		&			&			&				\\
$\infty$    	&		&	$u_{\rm tg}=$5.9170	&		&	 7.527(29)  	\\
\midrule
12	&	7.46910	&	7.232(11)	&	9.570(17)  	&	9.570(26)		\\
18	&	7.77340	&	7.220(11)	&	9.726(31)      	&	9.758(31)		\\
24	&	8.00000	&	7.2322(83)	&	9.734(36)      	&	9.732(37)		\\
	&		&			&			&				\\
$\infty$   	&	  	&	$u_{\rm tg}=$7.2325	&		&	 9.836(39)  	\\
\midrule
12	&	7.27130	&	7.943(12)	&	10.933(21)    	&	10.933(31)		\\
18	&	7.57350	&	7.927(12)	&	11.163(41)      &	11.202(41)		\\
24	&	7.80000	&	7.9290(95)	&	11.210(47)     	&	11.257(48)		\\
	&		&			&			&				\\
$\infty$   	&	  	&	$u_{\rm tg}=$7.9431	&		&	 11.386(51)  	\\
\midrule
12	&	7.07390	&	8.832(15)	&	12.975(30)      &	12.975(44)		\\
18	&	7.37430	&	8.826(15)	&	13.185(52)      &	13.223(54)		\\
24	&	7.60000	&	8.846(11)	&	13.370(68)    	&	13.339(73)		\\
	&		&			&			&				\\
$\infty$   	&	  	&	$u_{\rm tg}=$8.8322	&		&	 13.434(71) 	\\
\midrule
12	&	6.87710	&	9.994(18)	&	16.148(42)	&	16.148(63)		\\
18	&	7.17500	&	10.002(18)	&	16.596(61)	&	16.575(78)		\\
24	&	7.40000	&	9.999(14)	&	16.599(89)	&	16.585(96)		\\
	&		&			&			&				\\
$\infty$   	&	  	&	$u_{\rm tg}=$9.9945	&		&	 16.81(10) 	\\
\midrule
12	&	6.48810	&	13.947(35)	&	32.17(11)	&	32.01(21)		\\
18	&	6.77900	&	13.931(34)	&	33.29(20)	&	33.21(27)		\\
24	&	7.00000	&	13.916(27)	&	34.00(20)	&	34.00(25)		\\
	&		&			&			&				\\
$\infty$   	&	  	&$u_{\rm tg}=$13.91650	&		&	 34.52(29) 		\\
\bottomrule
\caption{Raw data for the lattice step scaling function $\Sigma(u)$ at the values of $u$ tuned for the $u$-by-$u$ continuum extrapolation - see sec.~\ref{s.ubyuf}.  The small mismatch in fixing the value of $u$ is corrected for
by slightly shifting the lattice step scaling functions to a constant value $u_{\text{tg}}$, leading to the values
$\Sigma(u_{\text{tg}})$ given in the table. We also give the continuum step scaling function obtained from 
the continuum extrapolations displayed in fig.~\ref{Figure_extrapl_fixed_u}.
}
\label{table_u_cont}
\end{longtable}

\section{Matching with the SF scheme}
\label{sec:matching-with-sf}

In order to match the values of the coupling $\lambda_{\rm TGF}(\mu)$
to the SF scheme, we use a set of simulations performed at the same
values of the bare coupling but on a symmetric lattice with size
$L_{\rm SF}=\tilde L/3$ with SF boundary conditions and an abelian
background field~\cite{Luscher:1992an}.  
The values of the SF coupling in these simulations can be checked in
reference~\cite{DallaBrida:2019wur}~\footnote{Reference ~\cite{DallaBrida:2019wur} gives the values of
the $SU(3)$ Yang-Mills coupling $\bar g^2_{\rm SF}$ related to the 't
Hooft coupling used in our work through 
the standard relation: $\lambda_{\rm SF} = 3 \bar g_{\rm SF}^2$.}.

We perform a fit, discarding the SF data with $\tL=18$,  of the form
\begin{equation}
  \frac{1}{\lambda_{\rm SF}(\mu)} - \frac{1}{u} = \sum_{n=0}^{N_l} c_k u^k +
  \left( \frac{1}{\tilde L^2}\right) \times
  \sum_{n=0}^{8} \rho_k u^k\,,
\end{equation}
where $c_k$ are the fit coefficients and
\begin{equation}
  u = \lambda_{\rm TGF}(\mu/0.9)\,.
\end{equation}

\begin{figure}[t]
  \centering
  \includegraphics[width=\textwidth]{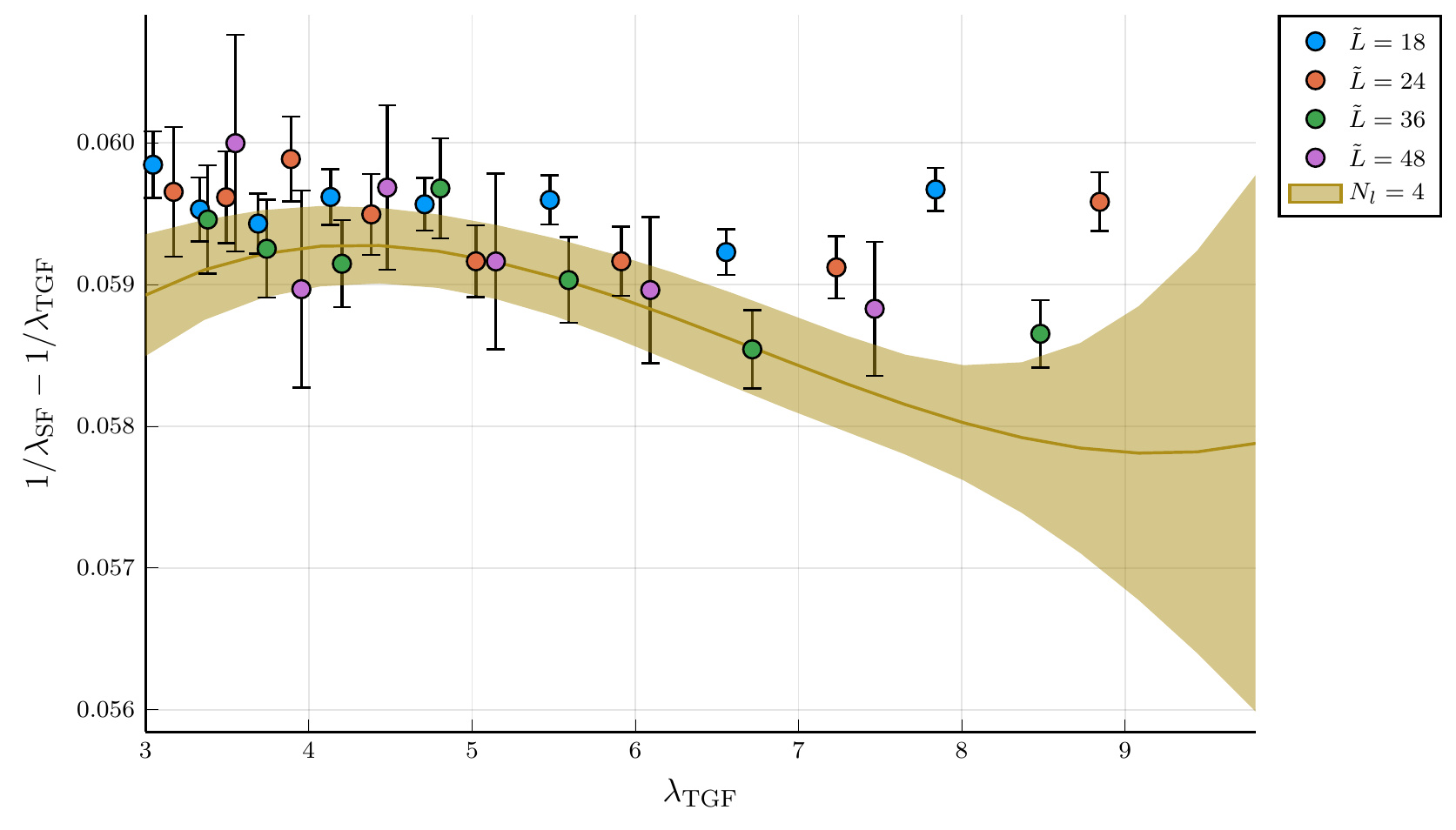}
  \caption{Non perturbative matching between the SF and the
    TGF couplings.
  We use the available data for the SF coupling in~\cite{DallaBrida:2019wur}.
In the fit we discard the data with $\tilde L=18$, but these are plotted to
show the consistency of the fit.}
  \label{fig:sf}
\end{figure}

The result of the fit can be seen in Figure~\ref{fig:sf}.
This fit can only be trusted for $3 < \lambda_{\rm TGF} < 6$, since
data on the SF scheme at higher energies is not available in the literature.

\section{Dilute gas approximation}
\label{sec:diluteg}
In this appendix we present the derivation of the dependence of the coupling on the topological charge in the dilute gas approximation.
The starting point is the continuum expression for the TGF coupling:
\be
\lambda (\mu) =
\frac{ 128 \pi^ 2 t^2 }{3 N  {\cal A}(\pi c^2)} \langle E\left(t\right) \rangle \Big |_{\sqrt{8t}=c \tl = \mu^{-1}} 
\, .
\ee
By setting ${\cal A}(x) = x^2 \widehat {\cal A} (x)$, and $ (8 t)^2 = c^4 \tilde l^4 = c^4 N^2 V$, 
with $V$ the volume of the asymmetric box,   one gets:
\be
\lambda (\mu) = \frac{ 2 N V }{  3 \widehat{{\cal A}}(\pi c^2)} \langle E(t) \rangle  \Big |_{\sqrt{8t}=c \tl = \mu^{-1}} \, .
\ee
We will use that:
\be
\langle  E(t) \rangle  = \frac{1}{2} \langle \Tr \left (G_{\mu \nu}^2 (x,t) \right) \rangle = \frac{1}{2 V} \left \langle \int d^4 x \, 
\Tr \left (G_{\mu \nu}^2 (x,t) \right) \right \rangle_Q
\, .
\ee
In  a configuration with well separated instantons and anti-instantons, the classical Euclidean action can 
be approximated by $S_0 (n+\bn)$, with $S_0$ the one-instanton action and $n(\bn)$ the number of 
instantons(anti-instantons). Therefore, one can compute the contribution to the expectation value of 
the energy density at leading order in this approximation by estimating the average number of instantons 
plus anti-instantons in the flowed ensemble:
\be
\langle  E(t)  \rangle^{\rm dig} \sim \frac{S_0}{V} \, \langle n +\bn \rangle \, .
\ee
Here we rely on the assumption that the flow removes short distance fluctuations but
preserves instantons, note though that this is not the case for instanton-anti-instanton pairs which
will only be preserved at distances larger than $\sqrt{8 t}=c \tl$. 

Finally, the evaluation of the coupling can be restricted to sectors of definite topological charge $Q$, cf. eq.~\eqref{eq:obsQ}, leading to:
\be
\lambda_Q^{\rm dig} (c) \sim \frac{2 N S_0}{3 \widehat {\cal A}(\pi c^2) } \, \langle n +\bn \rangle_Q \, .
\ee

In the remaining of this section we will determine the expectation value of $n+\bn$ considering
two different scenarios, a dilute gas of ordinary instantons with $S_0=8\pi^2$, and one made of 
fractional instantons of action $S_0=8\pi^2/N$.

\subsection{Ordinary instantons}

The basic asumption in the dilute gas approximation is that the gas is made of objects with 
positive (+1)  and negative (-1) topological charge that are otherwise equally probable; the probality 
to create either of them per unit volume is given by $R$. Under this hypothesis, it is easy to determine
the partition function restricted to the sector of topological charge $Q$, it is given by:
\be
Z_Q = {\cal C} \sum_{n,\bn} \frac{1}{n! \bn!}  ( RV)^{n +\bn} \delta(n-\bn -Q) =
{\cal C} \int_0^{2\pi} \frac{d\theta}{2\pi} e^{-iQ\theta} \exp \{ 2 RV \cos \theta\} = {\cal C} I_Q(2RV)
\, ,
\ee
where we have used the integral representation of the $\delta$ function:
\be
\delta(x)= \frac{1}{2\pi} \int_0^{2\pi} d\theta e^{i\theta x}
\, ,
\ee
and where $I_Q(x)$ stands for the modified Bessel function of the first kind.
It is easy to evaluate $\langle n +\bn \rangle_Q$ in this approximation, it is given by:
\be
\langle  n + \bn  \rangle_Q = \frac{\partial \log I_Q(x)}{\partial \log x} =
x  \left (\frac{I_{Q-1}(x) + I_{Q+1}(x)}{2 I_Q(x)} \right) =  Q + x \, \frac{I_{Q+1}(x)}{I_Q(x)}
\, ,
\label{eq:ediluteg}
\ee
where $x=2RV$.

We can also compute the topological susceptibility in this approximation. The partition function in the $\theta$ vacuum is given by:
\be
Z(\theta) =   {\cal C} \sum_{n,\bn} \frac{1}{n! \bn!}  ( RV)^{n +\bn} e^{i (n - \bn) \theta}
= {\cal C}  e^{2 RV \cos\theta} \, ,
\ee
and the topological susceptibility can be computed from:
\be
\chi = \left \langle \frac{(n - \bn)^2}{V} \right \rangle = - \frac{1}{V Z(\theta)} \frac{\partial^2Z(\theta)}{\partial^2 \theta}  \Big |_{\theta=0}= 2R
\, .
\ee
In the dilute gas approximation $R$ is determined from the determinant of the fluctuation operator around one instanton. The result at one-loop order in infinite volume reads
\be
R =  A \, \left(\lambda(\mu)\right)^{-\frac{n_{c}}{2}} \,  e^{-N S_0/ \lambda(\mu)}  \int_0^{\infty} \frac{d\rho}{\rho^5} \, (\rho \mu)^{2 N b_0S_0}
\, ,
\ee
where $n_{c}$ is the number of collective coordinates, equalt to $4N$ for ordinary $SU(N)$ instantons. The formula up to two loops including the prefactor has been computed in ref.~\cite{Morris:1984zi}. The resulting expression is infrared divergent, 
although on a finite volume it would be cutoff by the size of the box.
Note however, that for the determination of $\lambda_Q^{\rm dig}$ we only need the relation $x=2RV=\chi V$,
where one can use the measured topological susceptibility to determine the input parameter $x$ and evaluate the coupling using eq.~\eqref{eq:ediluteg}.   

\subsection{Fractional instantons}

In the case of fractional instantons, the only difference stems from the fact that instantons contribute with charge $\pm 1/N$ to the partition function. 
Note that the twist used in our simulations is in the class of the so-call orthogonal twists, 
satisfying $n_{\mu \nu} \tilde n_{\mu \nu} /4 = 0$ (mod $N$) and therefore the total topological charge is still quantized in integer units. 
This fact has to be taken into account when formulating the dilute gas partition function for fractional instantons and therefore:
\be
Z_Q = {\cal C}  \sum_{n,\bn} \frac{1}{n! \bn!}  ( RV)^{n +\bn} \delta(n-\bn -N Q)
\, .
\ee
We can now use that:
\be
I_\alpha (x) = \sum_n \frac{1}{n! \Gamma(n+\alpha+1)} \left (\frac{x}{2} \right)^{2 n + \alpha}
\, ,
\ee
to obtain ($x=2RV$):
\be
Z_Q = {\cal C} I_{NQ} (x) \, ,
\ee
and 
\be
\langle  n + \bn   \rangle_Q =  x  \, \left (\frac{I_{NQ-1}(x) + I_{NQ+1}(x)}{2 I_{NQ}(x)} \right ) =
NQ + x \, \frac{ I_{NQ+1}(x)}{I_{NQ}(x)}
\, .
\ee

As for the topological susceptibility, using:
\be
Z(\theta) = \sum_{Q\in \mathbf{Z}}  e^{i Q \theta}  Z_Q
\, ,
\ee
and the following representation for the Kronecker delta function:
\be
\delta (n-\bn -NQ) = \frac{1}{N} \sum_{k=1}^{N} e^{2 \pi i \frac{k}{N} (n-\bn)}
\, ,
\ee
one arrives at:
\be
Z(\theta) = \frac{{\cal  C }}{N} \sum_{k=1}^{N} \exp\left \{ x \cos \left (\frac{\theta+ 2 \pi k}{N}\right ) \right \}
\, ,
\ee
leading to the following expression for the $N=3$ topological susceptibility:
\be
\chi = \frac{x}{18V} \, \left ( 2 - \frac{3 (2 + x)}{2 +  e^{3 x/2}} \right ) \, .
\, .
\ee
The relation between $x=2RV$ and $\chi V$ is more involved in this case than for ordinary instantons,
 but nevertheless it can be numerically inverted to obtain $x$ in terms of $\chi V$.

In the dilute gas approximation, see i.e.~\cite{GarciaPerez:1993ab,GarciaPerez:1993jw}:
\be
R =  \widehat C \, \left (\lambda(\mu)\right)^{-2} \,  e^{- N S_0 / \lambda(\mu)}   \frac{1}{\rho^4} \, (\rho \mu)^{2 N b_0 S_0}
\, ,
\ee
where $\rho$ is the characteristic fractional instanton size. The peculiarity of fractional instantons vs ordinary ones is that the
collective coordinates only include the instanton location; in particular the size is not a moduli parameter and on a finite volume it is fixed
in terms of the volume of the box. 
Therefore, we expect $\rho \propto \tl$ in the small volume regime and $ \rho \propto 1/\Lambda_{QCD}$ for large volumes. 
When $\rho \propto \tl$, and setting $\mu=1/(c \tl)$, we can write:
\be
x=2RV =   C \frac{V}{(c\tl)^4} \, \left(\lambda(\mu)\right)^{-2} \,  e^{- 8\pi^2 / \lambda(\mu)} \Big |_{\mu=(c \tl)^{-1}} 
\, ,
\ee
and determine $x$, up to an overall normalization constant, from the measurement of the TGF coupling.


\clearpage
\addcontentsline{toc}{section}{References}
\bibliography{/home/alberto/docs/bib/math,/home/alberto/docs/bib/campos,/home/alberto/docs/bib/fisica,/home/alberto/docs/bib/computing}

\end{document}